\documentclass[sn-mathphys-num]{sn-jnl}

\usepackage[utf8]{inputenc}

\usepackage{graphicx}%
\usepackage{multirow}%
\usepackage{amsmath,amssymb,amsfonts}%
\usepackage{amsthm}%
\usepackage{mathrsfs}%
\usepackage[title]{appendix}%
\usepackage{xcolor}%
\usepackage{textcomp}%
\usepackage{manyfoot}%
\usepackage{booktabs}%
\usepackage{algorithm}%
\usepackage{algorithmicx}%
\usepackage{algpseudocode}%
\usepackage{listings}%
\usepackage{csquotes}
\usepackage[inline]{enumitem}
\usepackage[capitalise]{cleveref}
\usepackage{todonotes}
\usepackage{subcaption}
\usepackage{tabularx}
\usepackage{xcolor}
\usepackage{changepage}
\usepackage{ltablex}
\usepackage{catchfile}

\newlist{questions}{enumerate}{2}
\setlist[questions,1]{label=RQ \arabic*:,ref=\arabic*,left=0em}
\setlist[questions,2]{label=\thequestionsi.\arabic*,ref=\thequestionsi.\arabic*,left=0em}
\crefname{questionsi}{RQ}{RQs}
\Crefname{questionsi}{RQ}{RQs}
\crefname{questionsii}{RQ}{RQs}
\Crefname{questionsii}{RQ}{RQs}

\newcommand*{\rquestion}[2]{%
    \expandafter\gdef\csname item#1\endcsname{#2}%
    \label{#1}#2%
}
\newcommand*{\rqtext}[1]{\csname item#1\endcsname}

\newlist{ic}{enumerate}{2}
\setlist[ic,1]{label=IC \arabic*:,ref=\arabic*,left=0em}
\setlist[ic,2]{label=\theici.\arabic*,ref=\theici.\arabic*,left=0em}
\Crefname{ici}{IC}{ICs}
\Crefname{icsii}{IC}{ICs}
\crefname{ici}{IC}{ICs}
\crefname{icsii}{IC}{ICs}

\newlist{ec}{enumerate}{2}
\setlist[ec,1]{label=EC \arabic*:,ref=\arabic*,left=0em}
\setlist[ec,2]{label=\theeci.\arabic*,ref=\theeci.\arabic*,left=0em}
\Crefname{eci}{EC}{ECs}
\Crefname{ecii}{EC}{ECs}
\crefname{eci}{EC}{ECs}
\crefname{ecii}{EC}{ECs}

\newcommand{\quotebox}[1]
{
  \begin{center}
    \fcolorbox{white}{blue!15!gray!15}{
      \begin{minipage}{.9\linewidth}\vspace{10pt}
        \center
        \begin{minipage}{\linewidth}{\space}{\setlength{\parindent}{1.5em}\noindent\small#1}{\hspace{1.5em}\break\null}
        \end{minipage}
      \end{minipage}
    }
\end{center}
}

\newif\ifshowappendix
\showappendixtrue
\showappendixfalse
\newcommand{\appendixtable}[1]{\ifshowappendix\ (see also \cref{#1})\fi}

\usetikzlibrary{svg.path}
\definecolor{orcidlogocol}{HTML}{A6CE39}
\tikzset{
  orcidlogo/.pic={
    \fill[orcidlogocol] svg{M256,128c0,70.7-57.3,128-128,128C57.3,256,0,198.7,0,128C0,57.3,57.3,0,128,0C198.7,0,256,57.3,256,128z};
    \fill[white] svg{M86.3,186.2H70.9V79.1h15.4v48.4V186.2z}
                 svg{M108.9,79.1h41.6c39.6,0,57,28.3,57,53.6c0,27.5-21.5,53.6-56.8,53.6h-41.8V79.1z M124.3,172.4h24.5c34.9,0,42.9-26.5,42.9-39.7c0-21.5-13.7-39.7-43.7-39.7h-23.7V172.4z}
                 svg{M88.7,56.8c0,5.5-4.5,10.1-10.1,10.1c-5.6,0-10.1-4.6-10.1-10.1c0-5.6,4.5-10.1,10.1-10.1C84.2,46.7,88.7,51.3,88.7,56.8z};
  }
}
\renewcommand{\orcidlogo}{\tikz[yscale=-1,transform shape]{\pic{orcidlogo}}}
\renewcommand{\orcid}[1]{%
\resizebox{8px}{8px}{%
      \href{https://orcid.org/#1}{\orcidlogo}}%
}

\begin{document}

\title[Application of AI to formal methods]{Application of AI to formal methods --- an analysis of current trends}


\author[1]{\fnm{Sebastian} \sur{Stock} \orcid{0000-0002-2231-8656}}\email{sebastian.stock@jku.at}

\author*[2]{\fnm{Jannik} \sur{Dunkelau} \orcid{0000-0003-0819-5554}}\email{jannik.dunkelau@hhu.de}

\author[1]{\fnm{Atif} \sur{Mashkoor} \orcid{0000-0003-1210-5953}}\email{atif.mashkoor@partner.jku.at}

\affil[1]{
    \orgdiv{Institute of Software System Engineering},
    \orgname{Johannes Kepler University Linz},
    \orgaddress{
        \street{Altenbergerstraße 69},
        \city{Linz},
        \postcode{4040},
        \country{Austria}}}

\affil*[2]{
    \orgdiv{Heinrich Heine University Düsseldorf},
    \orgname{Faculty of Mathematics and Natural Sciences},
    \orgaddress{
        \street{Universitätsstraße 1},
        \city{Düsseldorf},
        \postcode{40225},
        \country{Germany}}}

%


\abstract{%
    \textbf{Context:}
    With artificial intelligence (AI) being well established within the daily lives of research communities, we turn our gaze toward formal methods (FM). FM aim to provide sound and verifiable reasoning about problems in computer science.

    \textbf{Objective:}
    We conduct a systematic mapping study to overview the current landscape of research publications that apply AI to FM\@.
    We aim to identify how FM can benefit from AI techniques and highlight areas for further research.
    Our focus lies on the previous five years (2019--2023) of research.

    \textbf{Method:}
    Following the proposed guidelines for systematic mapping studies, we searched for relevant publications in four major databases, defined inclusion and exclusion criteria, and applied extensive snowballing to uncover potential additional sources.

    \textbf{Results:}
    This investigation results in 189 entries
    which we explored to find current trends
    and highlight research gaps.
    We find a strong focus on AI in the area of theorem proving
    while other subfields of FM are less represented.

    \textbf{Conclusions:}
    The mapping study provides a quantitative overview of the modern state
    of AI application in FM\@.
    The current trend of the field is yet to mature.
    Many primary studies focus on practical application, yet we identify a lack of theoretical groundwork,
    standard benchmarks, or case studies.
    Further, we identify issues regarding shared training data sets and
    standard benchmarks.

}




\maketitle

\section{Introduction}%
\label{sec:introduction}
Artificial intelligence (AI),
and especially the subdomain of machine learning (ML),
is becoming increasingly relevant for all industries
and increasingly penetrates research communities,
as several studies show~\cite{Lu2019,Elahi2023,Jiang2023,Chang2023}.
Consequently,
AI applications also gained popularity in software development
~\cite{barenkamp2020applications,shafiq2021literature}.

This study asks whether this trend is also observable within the formal methods communities.
Formal methods (FM) correspond to a mathematically rigorous approach to software and
systems development in which the concern is to provide a
formal representation of hardware, software, or systems engineering
problem~\cite{Abran2004}.
This representation then lends itself to mathematical assessments such as
correctness, soundness, or well-definedness.
These assessments can be easily reproduced and followed by the practitioner.

However, the needed guarantees and the rigorousness of these methods seem to conflict with AI applications, which often show nondeterministic behavior and mostly correspond to black-box techniques. The trade-off appears obvious: Injecting AI's nondeterminism and unpredictability into FM's rigorousness could lead to faster results and proofs with less human intervention at the price of fundamental guarantees. This point of tension seems interesting and needs further investigation, but it is not the only way how AI can enrich FM\@.
Much like in software development, AI-driven tools might still supplement the formal development process
without impeding the correctness of the results. From personal experience, we noticed increased publications applying AI techniques to FM in recent years, suggesting that the FM field also follows the overall trend.

To substantiate this and to obtain an overview of which FM subdomains already utilize AI, we performed this systematic mapping study (SMS)~\cite{petersen2015guidelines}. The primary goal is to observe the development of AI applications quantitatively
to the field of FM\@.

In this work, we report on our journey and the results of our SMS,
accessing the quantitative nature of the topic of AI applications in the FM domain.
This represents a typical goal for this kind of study,
as pointed out by \citet{Kitchenham2010}.
Mainly, we are interested in which FM domains are targeted by AI improvement already, which AI or ML applications have been investigated in the literature so far,
where potential research gaps exist,
and whether there appears to be consensus within the FM domains of which
AI techniques seem most beneficial.
A challenge in this assessment is the wide variety of topics on both sides.
AI and FM are huge topics, and their potential overlap may be considerable.
This study aims to be the first assessment of the general research landscape,
i.e., we assume a mainly quantitative lens on the topic.
Consequently, we highlight areas that received more attention in the literature,
as well as those that received less attention.
However, we refrain from doing a qualitative assessment,
e.g., a systematic review of the literature.
Our goal is to gain an initial overview of the field's landscape.

While our search process resulted in a data set of 457 highly relevant publications covering more than 50 years of research applying AI to FM, we investigated only articles from the last five years in more detail.
This period is chosen as it corresponds to a peak in the number of publications
as indicated by previously mentioned studies
and also allows us to focus on the most recent developments.
Consequently, we have 191 primary studies, two of which are pure data set presentations.
We analyzed the remaining 189 studies to determine
which AI techniques were found to be applicable in which FM areas.
Further, we derive research suggestions from our insights
by pointing out research gaps that might further mature the field.
Both the data set of the last five years and the complete data set of
457 publications are made publicly available to interested researchers
at \url{https://github.com/hhu-stups/ai4fm-studies}.

The rest of this mapping study is structured as follows:
\Cref{sec:background} introduces the relevant subdomains of FM and AI\@.
It will also provide an overview of our understanding of AI\@.
We present our leading research questions in \cref{sec:goal}
and detail our search and selection process in \cref{sec:strategy}.
\Cref{sec:result} presents the results of the search while
\Cref{sec:discussion} engages in a deeper discussion about these results.
In \cref{sec:threats}, we discuss potential threats to the validity
of our conducted systematic mapping study.
\Cref{sec:related_work} presents related studies we found during our research
and compares them against our efforts.
Finally, we conclude in \cref{sec:conclusion}.

\section{Background}%
\label{sec:background}

\subsection{Systematic mapping studies}
Systematic mapping studies (SMS), as proposed by Kitchenham et al.~\cite{Kitchenham2007}, aim to assess the quantitative nature of a research field. Here, the goal is to give an overview of the available research. The counterpart to an SMS would be a systematic literature review where the main goal is to assess the quality of contributions.

SMS are typically performed using a multi-step approach. First, the field of interest is selected. Then, research questions are formulated, and inclusion criteria (IC) and exclusion criteria (EC) are defined. IC and EC serve as filters over the collected primary studies to help select relevant results.

After that, the search query is crafted. The goal is to apply the search query to the scientific meta-search engines to retrieve a good corpus of primary studies that answer the research questions. With the retrieved corpus, the IC and EC are applied. If indicated, snowballing is done, usually until a stable closure of the corpus is reached. With a corpus constructed, the aim is to answer the research questions.

\subsection{Formal methods}%
\label{sec:background-fm}

Formal methods (FM) describe a mathematically rigorous approach to
design and assess software as well as hardware systems,
and are concerned with analysis, validation, and verification
at any part of the respective system's
life cycle~\cite{woodcock2009formal,clarke1996formal}.
Employing FM allows
to formulate precise statements of desired functionality or requirements
in form of a \emph{formal model} (or \emph{formal specification})
while not constraining the possible implementation thereof.
The formal model is represented in a mathematically approachable form
which lends itself to reasoning and hence allows
rigorous analysis of critical properties such as
correctness, safety, soundness, or well-definedness.
The following section overviews FM's creation and this work's most relevant reasoning methods.

\paragraph*{Model checking.}
    Model checking~\cite{Baier2008, Clarke2018} explores a program's or system's state space. The state space sets all possible value constellations that the program can achieve. In its most basic form, model checking aims to explore all reachable states and check if they are faulty, i.e., violate any specified properties. If this is the case, a counterexample is found.
    Otherwise, the system works correctly, as no faulty states are
    reachable.

    There are different types of model checking, besides classical, explicit state model checking.
    Noteworthy for this study are
    linear temporal logic (LTL) model checking,
    symbolic model checking,
    and statistical model checking.
    While explicitly visiting states, LTL model checking focuses on temporal properties encoded in LTL formulas, which are checked against execution traces of the state space.
    Symbolic model checking abstracts the state space and makes a symbolic
    evaluation of a complex expression; thus, it shrinks the overall state space for the price of precision. With enough abstraction, it can handle even infinite state spaces
    for the price of precision.
    Statistical model checking investigates state spaces created by assigning
    probabilities on transitions between states and provide
    probabilistic guarantees.

\paragraph*{Theorem proving.}
    A core application of theorem proving (TP)
    is to evaluate statements about a formal model's
    internal consistency and behavioral integrity.
    When talking about theorem proving, we usually assume some formal representation of the problem
    for which proofs are formulated and discharged.
    Discharging proofs can either be done by a fully automatic theorem prover (ATP)~\cite{Gallier2015} that finds a solution on its own,
    or an interactive theorem prover (ITP)~\cite{Bertot2013}
    that requires human input for individual proof steps and transformations.
    ITPs provide a user interface to keep track of progress, sub-goals,
    and available properties.

\paragraph*{SAT/SMT solving.}
    The Boolean satisfiability problem (SAT)~\cite{Biere2009}
    refers to finding the right model for a Boolean formula to satisfy the formula.
    Historically,
    SAT was the first problem found to be NP-complete
    and has since drawn a broad research interest.

    Satisfiability modulo theories (SMT)~\cite{Barrett2018} is a more generalized form of the SAT, which enriches the problem setting with various theories such as arithmetic, data structures, or set theory.

    In the context of FM,
    SAT and SMT solving
    find applications in theorem proving~\cite{Brown2013}, bounded model checking~\cite{Shtrichman2000},
    equivalence checking~\cite{Goldberg2001} or test generation~\cite{Zeng2005}.

\paragraph*{Synthesis.}
    Under the term of synthesis, we group all attempts automatically or semi-automatically
    create (parts of) formal models during the formal development process,
    such as the generation of models from natural language specification or vice versa.
    We further group the idea of automated model repair under this aspect,
    where, for a model with some violation, a fix is synthesized that
    rectifies the violation.

\paragraph*{Other categories.}
    In the scope of this mapping study, we may encounter FM approaches and techniques that do not belong to any previous group.
    We will group these as \emph{other} FM techniques.
    For instance,
    this includes meta-approaches such as selecting the right verification algorithm,
    general program analysis, or termination analysis.

\subsection{Artificial intelligence}%
\label{sec:background-ai}

Artificial Intelligence (AI) is a catch-all phrase for various
techniques and approaches that aim to imitate seemingly \emph{intelligent} behavior.
\citet{Simmons1988} define the term as \textquote{[\ldots] behavior of a machine which,
    if a human behaves in the same way, is considered intelligent}.
They further emphasize the problem with the term itself,
pointing out the difficulty of precisely defining what
it means for a human to be intelligent.
Nevertheless, AI is regularly used in daily life and the research community.

An important AI paradigm is machine learning (ML).
The goal of ML algorithms
is to observe and extract patterns within the data
which can then be applied to the automation of a given task
~\cite{mitchell1997machine}.
Instead of manually defining a set of rules for pattern recognition, machine learning algorithms uncover such rules automatically, a process referred to as \emph{learning}.
ML algorithms experience the given data in some manner
and improve their exhibited performance to solve the task at hand gradually, the more experience they gather.

Below, we list and briefly explain AI and ML algorithms
which are most relevant to this mapping study.
For the reader familiar with AI, the selection seems mostly ML rather than AI. Indeed, we aimed to find applications of AI to FM, and the search query we will discuss later was prepared accordingly. However, to bring a result up front, most contributions use some form of ML.

\paragraph*{Neural networks and deep learning.}
    Neural networks (NNs)~\cite{rosenblatt1962principles} build the foundation of what is known today as the area of
    deep learning~\cite{goodfellow2016deep}.
    A deep neural network (DNN)
    consists of a layer of inputs, a layer of outputs,
    and one or more hidden layers.
    Inputs are forwarded layer-wise and combined first by a weighted sum
    and a follow-up non-linear transformation.
    The clue is the training of a DNN, which starts the network with randomly initialized weights for the summations above, but adjusts them gradually in the learning process to converge to the
    desired function.

    A considerable benefit of neural networks lies in their variety.
    Different approaches and architectures exist for deep learning
    in specialized environments with differently shaped data.
    The most important for this work are convolutional neural networks (CNNs)~\cite{lecun2010convolutional} for images,
    recurrent neural networks (RNNs)~\cite{jordan1997serial} for sequential data,
    the transformer architecture~\cite{Vaswani2017} for language processing,
    and graph neural networks (GNNs)~\cite{Zhou2020}.
    Today, deep learning is considered one of the most popular topics in ML~\cite{sarker2021deep}.

\paragraph*{Reinforcement learning.}
    Reinforcement learning (RL)~\cite{Kaelbling1996} is foremost a machine learning paradigm.
    Given an environment, a set of actions that change the environment, and a reward function that quantifies a given environmental
    state, the RL agent aims to learn a policy over their available actions that maximizes the cumulative reward over time.
    Initially, the agent does not know the environment or the task it is supposed to solve, but can only learn from feedback through the received
    rewards~\cite{sutton2018reinforcement}.

\paragraph*{Natural language processing.}
    Natural language processing (NLP)~\cite{Chowdhary2020}
    describes the research area of, as the name suggests,
    processing and evaluating texts of human language.
    This includes
    information extraction,
    language translation tasks,
    text classification,
    semantic analysis,
    and
    natural language generation and dialogue systems~\cite{Chowdhary2020,Khurana2023}.

    One of the most recent developments is large language models (LLMs)~\cite{Chang2024}
    which gained significant popularity and even worldwide attention
    with the release of ChatGPT~\cite{Wu2023chatgpt}. For the sake of compactness, we count contributions with LLMs towards NLP.

\paragraph*{Genetic/evolutionary algorithms.}
    Genetic and evolutionary algorithms (EAs)~\cite{mitchell1998introduction} describe a family of optimization algorithms rooted in evolution.  An EA starts with a randomly assembled population of candidates for a given optimization problem and a fitness function. The fitness function measures how well a candidate solves the problem.
    Parent candidates are chosen via a random selection process that is nonetheless influenced by
    each individual's fitness. These are adapted using mutation or
    recombination to produce a new child population.
    Repeating this process converges the individuals toward suitable solutions
    for the problem at hand.

\paragraph*{Statistical approaches.}
    In contrast to the abovementioned techniques, we consider techniques from more classical, statistical ML\@.
    These differ from the above methods as they do not depend on deep learning
    and are typically faster to compute in comparison.
    This does, however, not imply that they are not competitive.
    The considered statistical approaches include
    support vector machines (SVM)~\cite{Kecman2005},
    logistic and linear regression (LR)~\cite{montgomery2021introduction},
    k nearest neighbors (KNN)~\cite{cover1967nearest},
    decision trees (DT)~\cite{breiman1984classification}
    and random forests (RF)~\cite{breiman2001random},
    or Bayesian inference approaches~\cite{Tipping2004}.

\subsection{Contribution classification}%
\label{sec:contribution-classification}
We classified the contributions into various types. The aim is to show the different approaches to target the same research area.
The different types of contributions are distinguished as follows:

\paragraph*{Tool.} Contributions that provide practical means to solve an FM problem,
        for instance, in the form of a binary or source code release.
\paragraph*{Tool enhancement.} Contributions that aim to enhance or complement an existing tool.
\paragraph*{Approach.} Contributions that propose an overall style or idea to overcome a problem. An approach is a generalized concept that details overcoming and resolving a problem. The approach remains more theoretical and does not involve tested or empirically proven steps.
\paragraph*{Methodology.} Contributions that realize and test an approach.
\paragraph*{Framework.} Contributions that propose a conceptual structure intended to support or guide the construction or expansion of a solution for an actual problem.
\paragraph*{Case study.} Contributions apply an approach, a methodology, or a framework to a real-world problem.
\paragraph*{Benchmark.} Contributions that test an existing tool or methodology.
\paragraph*{Idea.} Contributions that focus on a brief overview or discussion of an idea to solve a problem. Compared to an approach, it does not lay out larger theories but serves as a quick pitch.
\paragraph*{Training data.} An openly accessible data set that claims to be reusable outside of the replication of the contribution it was originally used in.

\subsection{Publication venue classification}%
We further divide the publication venues into the following categories
to better assess the respective maturity of the contributions:
\paragraph*{Workshop and short papers.} We grouped contributions in this category if they are
        \begin{enumerate*}[label=(\alph*)]
            \item rather short (4 pages or less) or
            \item if the venue they appeared in has workshop character, i.e., is classified as a workshop, calls itself a workshop, or focuses on the in-person presentation of results but requires submission of an extended abstract.
        \end{enumerate*}
\paragraph*{Conference proceedings.} Here, we group all the contributions that are published in the proceedings of a conference. From a full conference paper, we expect a more in-depth explanation of the result than compared to a workshop or short paper, ultimately resulting in an overall more sizable contribution.
\paragraph*{Journal articles.}  Here, we group all contributions published as part of a journal volume. As journal articles tend to consist of more pages and go through a more thorough reviewing process, the results of journal articles are expected to have a broader scientific basis for their claims.
\paragraph*{Book chapters.} As the name suggests, these contributions are part of a larger collection of articles bundled in one book. As the production of scientific books takes a significant amount of time, the reader expects a high quality of the provided scientific contribution that is considered, by that time, state of the art.

\section{Research questions}%
\label{sec:goal}\label{sec:rqs}

To get an overview of the field,
we divide our research questions (RQs) into two main areas.
First,
\cref{rq:timeline,rq:pub-types,rq:contributions} focus on the overall demographics of the research field.
With these questions, we want a basic quantitative overview
to explore the field's maturity.
Second,
\cref{rq:ai-types,rq:fm-types,rq:ml-to-fm-dist,rq:datasets} focus on the content of the contributions and aim to map out the field, highlighting potential gaps and showcasing quantitative interest in respective subtopics.
Our research questions are defined as follows.

\begin{questions}
    \item \rquestion{rq:demographics}{Demographics of the research area}
    \begin{questions}
        \item \rquestion{rq:timeline}{What is the research publication timeline, and is there a trend?} \\
        \emph{Rationale:} We are interested in knowing whether the field experiences a growth or decline in interest,
            or if no trends are detectable at all.
        \item\rquestion{rq:pub-types}{Which publication venues are most frequent in the field?}\\
        \emph{Rationale:} We want to explore the field's maturity, whereby books, book chapters, and journal articles indicate more mature results.
        \item\rquestion{rq:contributions}{What are the main contributions provided by the primary studies?} \\
        \emph{Rationale:} We want to learn where the community focuses its efforts, e.g., more on the theoretical groundwork or practical application.
    \end{questions}

    \item\rquestion{rq:quality}{Content type of contributions}
    \begin{questions}
        \item\rquestion{rq:ai-types}{Which AI techniques and tools were used?} Are there any prevalent choices within the community?\\
        \emph{Rationale:} We want to know if there is an indication that some AI techniques are more relevant (or at least popular) than others.
        \item\rquestion{rq:fm-types}{What are the application areas of AI in FM\@?} Are there any commonly found FM techniques or tools?   \emph{Rationale:} We are curious about which FM techniques currently get the most attention or if there are dark spots in the literature.
        \item\rquestion{rq:ml-to-fm-dist}{What is the distribution of AI types in the different FM application areas?} \\
        \emph{Rationale:} Here we want to overlap the results of \cref{rq:ai-types,rq:fm-types}.
        \item\rquestion{rq:datasets}{Are the studies' employed data sets publicly available?}
        \emph{Rationale:} Especially in machine learning, having the source material available and accessible for experiments benefits the scientific community.
    \end{questions}

\end{questions}

\section{Search strategy}%
\label{sec:strategy}

\begin{figure}[ht]
    \centering
    \includegraphics{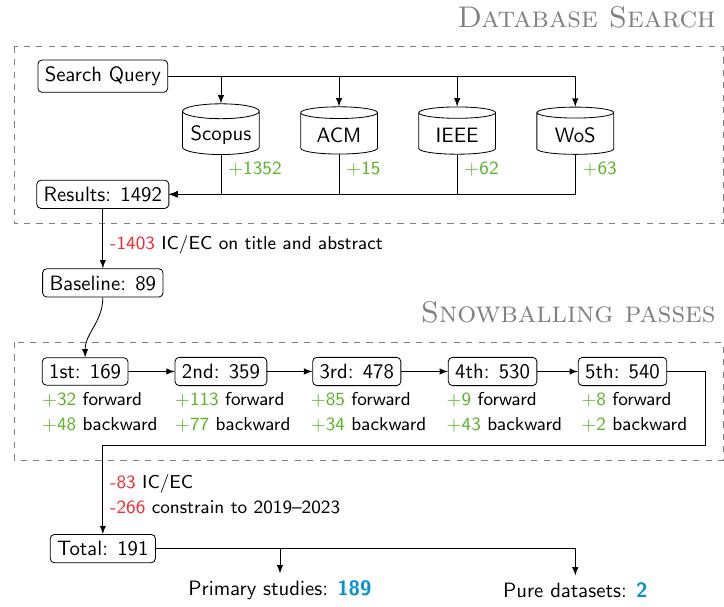}
    \caption{Visualization of the search process}%
    \label{fig:search_process}
\end{figure}

In the following, we explain our approach for aggregating the contributions for this mapping study
which we have visualized in \cref{fig:search_process}.
For this, we follow the suggested multi-step approach by~\citet{petersen2015guidelines},
which consists of an initial search in meta-search engines,
followed by a snowballing process.
As our selection of topics will show, it provided a challenging environment with overlapping search terms and related research areas.
We tweaked our search strategy to address these challenges and avoid a result set of tens of thousands of potential studies.

A vital role was played by the IC \& EC, which we applied twice. After the initial search, the first application was made to remove unrelated contributions that would introduce overheating into the snowballing. The second time was again after the snowballing to filter out wrongly selected contributions.

\subsection{Employed search query}%
\label{sec:search-query}
Finding a search query that would return a manageable amount of relevant contributions proved difficult.
The main problem was harnessing the bandwidth of terms used in the AI and FM domains.
Hereby, the difficulty lies in the often ambiguous use of terms like \enquote{formal method}, which can be used in the sense of formal methods as outlined in \cref{sec:background-fm} or as a term to describe an only somewhat formal approach to a problem. The situation is even more difficult for AI, as AI and ML are often used interchangeably. In contrast, essential publications may only use the specific name of an applied technique without naming the fields of AI or ML.

Another issue was the cross-pollution of our results with studies from the related research field of verification for AI and ML systems. This is essentially the other direction in which FM is applied to AI. This field is also a highly relevant topic and shares the majority of potential
search key phrases, further inflating our results tally.

As a reaction, we experimented with different approaches and probed the quantity and quality
of their respective results to find a suitable search query.
Highly abstract queries such as
\begin{center}
    \enquote{formal methods \& artificial intelligence}
\end{center}
produce multiple thousands of results
which we deemed impossible to process in a reasonable amount of time
and which contained many studies outside of our intended scope.
From probing some results as well as personal experience, we also knew that many relevant publications
tend not to use these high-level terms, but use nomenclature specific to their sub-community.
For instance, research regarding the formal B method
tends to simply use \enquote{B method} instead of \enquote{formal method}
as keyword.
On the other hand, we noticed how relevant publications also seemed to prefer to use the concrete
names of the applied AI algorithms, rather than stating they used AI in a
superficial manner.
Therefore, we were concerned that a search query that was too abstract would not be able to penetrate the relevant fields well enough.

As a result, we settled on a more complex query containing the precise terminology of specific algorithms.
For this,
we distinguish between two sets of terms,
the AI-terms (\cref{qu:ai})
and the FM-terms (\cref{qu:fm}).
Both correspond to a
disjunction of a set of selected terms from the respective field. The terms themselves were selected based on previous experiences in the field following the PICO (\textbf{P}opulation  \textbf{I}ntervention \textbf{C}omparision \textbf{O}utcome) approach
as suggested by \citet{Kitchenham2007}.

Here \textbf{P}opulation may refer to a specific field of FM or AI, such as formal model(ing) or deep learn(ing). \textbf{I}ntervention may refer to specific techniques or methodologies of FM or AI, such as SAT or SVM. \textbf{C}omparision was not applicable as we focused on a direct connection, while \textbf{O}utcomes shall contain at least strongly correlated AI and FM terms.

The final search query (\cref{qu:all}) now uses both term aggregates to produce a corpus of primary studies in a strict and targeted manner. We only looked for studies that use at least one AI term and at least one FM term in their titles, abstracts, and keywords. The reasoning is to get a good initial penetration of all key fields, while relevant but missed entries are found in the later search stages via snowballing.

\begin{figure}
    \begin{subfigure}{\textwidth}
        \quotebox{
            \textbf{FM-terms :=}
                \enquote{formal method}  OR
                \enquote{formal model}  OR
                \enquote{specification} OR
                \enquote{rigorous} OR
                \enquote{prove*}  OR
                \enquote{solv*}  OR
                \enquote{automated reason*}  OR
                \enquote{formal synth*}  OR
                \enquote{model check*}  OR
                \enquote{model repair} OR
                \enquote{premise selec*}  OR
                \enquote{SAT}  OR
                \enquote{SMT}  OR
                \enquote{smtlib}
        }
        \caption{FM keywords used in the query}%
        \label{qu:fm}
    \end{subfigure}
    \\ \bigskip

    \begin{subfigure}{\textwidth}
        \quotebox{
            \textbf{AI-terms :=}
                \enquote{artificial intelligence}  OR
                \enquote{AI}  OR
                \enquote{machine learn*}  OR
                \enquote{supervised learn*}  OR
                \enquote{unsupervised learn*}  OR
                \enquote{classification}  OR
                \enquote{regression}  OR
                \enquote{portfolio}  OR
                \enquote{deep learn*}  OR
                \enquote{neural net*}  OR
                \enquote{bayesian net*}  OR
                \enquote{reinforcement learn*}  OR
                \enquote{reinforcement agent}  OR
                \enquote{multi-armed bandit}  OR
                \enquote{knn}  OR
                \enquote{k nearest neighbours}  OR
                \enquote{support vector}  OR
                svm  OR
                \enquote{decision tree}  OR
                \enquote{random forest}  OR
                \enquote{gradient boost*}  OR
                \enquote{xgboost}  OR
                \enquote{logistic regress*}  OR
                \enquote{linear regress*}  OR
                \enquote{cluster*}  OR
                \enquote{computer vision} OR
                \enquote{nlp} OR
                \enquote{natural language proces*} OR
                \enquote{genetic prog*}  OR
                \enquote{genetic algorithm} OR
                \enquote{expert system}  OR
                \enquote{inductive logic progr*}
        }
        \caption{AI keywords used in the query}%
        \label{qu:ai}
    \end{subfigure}
    \\ \bigskip

    \begin{subfigure}{\textwidth}
        \centering
        \quotebox{TITLE(AI-terms) AND TITLE(FM-terms)
            AND ABSTRACT(AI-terms) AND ABSTRACT(FM-terms)
            AND KEYWORDS(AI-terms) AND KEYWORDS(FM-terms)}
        \caption{Unified search query}%
        \label{qu:all}
    \end{subfigure}
    \caption{Construction of the used search query.
        The query consists of a disjunction of search terms for AI and FM,
        respectively.
        The unified query (\protect\subref{qu:all}) enforces that at least one keyword
        from each subfield is present in the title, the abstract,
        and the keywords of a respective publication.
        An asterisk (*) indicates a wildcard character that can match any sub-word.
        For instance,
        \enquote{solv*} matches with \emph{solver} but also with \emph{solving}.
    }
\end{figure}

\subsection{Database search and processing}
The search was conducted in the last quarter of 2023. We used four meta-search engines\footnote{Springer's engine was not used because by the time of the search it performed extremely poor in dealing with large search queries, large result sets and filtering.} as suggested by~\citet{petersen2015guidelines}: IEEE\footnote{\url{https://ieeexplore.ieee.org/Xplore/home.jsp}}, Scopus~\footnote{\url{https://www.scopus.com/search/form.uri?display=basic\#basic}}, ACM\footnote{\url{https://dl.acm.org/}} and Web of Science (WoS)\footnote{\url{https://www.webofknowledge.com/}}.
We used the \emph{Guide to Computing Literature} for ACM. Furthermore, as the ACM search engine limits the number of wildcards, we split the search terms into subsets, performed the subsearches, and merged them into the required superset. For WoS, results seemed to differ depending on which institution had access. Therefore, we decided to take the more extensive result set. As far as the search engines allowed, we applied the EC, such as the area of publishing and the publication language. The result size after this step was 1492 entries. This number and all later numbers are without duplicates.

\subsection{Inclusion and exclusion criteria}%
\label{sec:ic-ec}
For the 1492 resulting contributions, we made two observations.
First, the amount was too large for a snowballing procedure to be feasible.
Second, while briefly looking at the corpus,
we discovered many contributions that should not have been selected,
i.e., contributions that included the term \enquote{specification} but meant it in a purely requirement engineering-minded context. Therefore, we decided to apply our IC and EC prematurely to the corpus as far as they applied, i.e., to the title and abstract. For this, we read the titles of our studies. If we could not decide whether the contributions should be included, we also conducted an abstract review, and very rarely, the contributions themselves were skimmed.

To achieve this for such a large corpus,
the first and second authors passed over all entries individually
and decided whether they should be included or excluded.
In agreement cases,
the respective studies were kept or discarded.
For disagreement, both authors discussed each instance together to reach an agreement.
The disagreement could not be settled this way in two cases, and the third author was consulted as a tiebreaker.
This procedure reduced the number of primary studies to 89, but took time.

The applied IC and EC 
are detailed in the following.
\begin{ic}
    \item\label{ic:1} The contribution focuses on applying techniques from the domain of AI to the domain of FM\@.
    \item\label{ic:2} The contribution is in English.
    \item\label{ic:3} The contribution has at least two pages of content (excluding references).
    \item\label{ic:4} If the contribution was submitted in different versions, we took longer. For instance, the journal version was taken for a conference contribution and subsequently published as an extended version. The journal version was taken for a journal-first submission and an invited version.
    \item\label{ic:5} The contribution is a data set used for ML\@.
\end{ic}

\begin{ec}
    \item\label{ec:1} The contribution appears not to have been reviewed in any form.
    \item\label{ec:2} The contribution is not available.
    \item\label{ec:3} The contribution is not posed in computer science.
    \item\label{ec:4} The contribution is unclear (even after reviewing the full text).
    \item\label{ec:5} The contribution solely focuses on solving a mathematical optimization problem.
    \item\label{ec:6} (During initial skim) The contribution focuses only on common or general constraint-solving techniques (without specializing in SMT/SAT) \label{ec:constraints}
    \item\label{ec:7} (During initial skim) The contribution focuses on portfolios for SAT solvers without explicitly mentioning any AI/ML technology  \label{ec:sat}
    \item\label{ec:8} (During initial skim) The contribution is not a primary study, i.e., it is another mapping study, survey, etc.  \label{ec:secondary}
    \item\label{ec:9} (Snowballing) The contribution is concerned with 2Sat, which is solvable in polynomial time.  \label{ec:sat2}
    \item\label{ec:10} (Snowballing) If the contribution proposes a way to synthesize some automaton (e.g., a Markov process), then the produced result must explicitly be used within the context of FM\@. This means that the synthesized automaton and its analysis with FM's help are described within the contribution.  \label{ec:sythesis}
\end{ec}

For the EC, some additional criteria were added in later steps in response to contributions encountered that were of low value for this study. For instance, \cref{ec:constraints} was added because many contributions focused on constraint solving, thus being closer to mathematics. \Cref{ec:sat} was added in response to a large influx of SAT contributions that mentioned \emph{learning} but focused on remembering already-seen clauses. \Cref{ec:secondary} is a rule that was introduced to not have secondary studies in the result set but to use those studies to ensure that even with a restrictive query, we cover as much of the field as possible. \Cref{ec:sythesis} was added in the fifth snowballing round, where we got many contributions that synthesized automatons, but no further analysis was mentioned. While an automaton may represent a formal model, analysis of larger automatons is very human-unfriendly compared to mathematical formulas or formal models written in a modeling language. Therefore, we expected the authors to subject the generated automaton to some validation, e.g., model checking or SAT/SMT solving, to enable reasoning about the validity of the generated automaton. \Cref{ec:sat2} was added as we found 2Sat as a problem irrelevant for this study, as solutions are already achievable in polynomial time.

\subsection{Snowballing}%
\label{sec:snowballing}
For the 89 entries, we conducted an extensive snowballing procedure~\cite{Wohlin2014} where we conducted both forward and backward snowballing.
Five iterations were needed until we eventually reached closure.
This extensive snowballing approach was meant to complement the strict search query
and to uncover relevant yet initially missed studies or even subfields of research.
Therefore, snowballing served as a means to alleviate these threats to validity.

For the backward snowballing, we consulted the references of a given contribution
and selected promising titles
or publications that were explicitly highlighted in the respective related work sections.
If a publication appeared promising, but the contribution was unclear from its title alone, the abstract was also consulted.
For the forward snowballing, we used Google Scholar and applied the same title and abstract procedure.
This brought the number of contributions up to 540.
After concluding the snowballing, we applied the IC and EC a second time, reducing the number of contributions to 457.

\subsection{Filtering by years}%
\label{sec:filter-years}
While the 457 results are all highly relevant, the number of contributions was too large to conduct a deeper investigation necessary to answer all the RQs that require a more in-depth consideration of the contributions, i.e., \cref{rq:pub-types,rq:contributions,rq:quality}. Therefore, we restricted the result tally to studies published between 2019 and 2023.
While this shifts our focus on the most recent developments in the field, it aligns with the peak in this period we discussed back in \cref{sec:introduction}.
By only considering studies from 2019 to 2023, we reduced the corpus to 191 primary studies.
Of these, 2 studies purely provide data sets for future AI applications in the field of FM
without showcasing any form of direct AI application themselves
(\cref{ic:5}).
That leaves us with a final tally of 189 primary studies
to conduct this mapping study.

\section{Results}%
\label{sec:result}
In the following section, we will discuss the results. For this, we will answer the individual research questions and aim to make a cross-analysis between individual research questions, thus extracting as much information as possible from the existing dataset. We refer to \Cref{sec:discussion} for a discussion of the observed results.
An overview of the results matched to their respective application domain in FM
can be found in
\cref{tab:sorted_primary_studies}.

\subsection{RQ~\ref{rq:demographics}: \rqtext{rq:demographics}}%
\label{subsec:quantity}

\subsubsection{RQ~\ref{rq:timeline}: \rqtext{rq:timeline}}
\begin{figure}[!h]
\centering
    \includegraphics{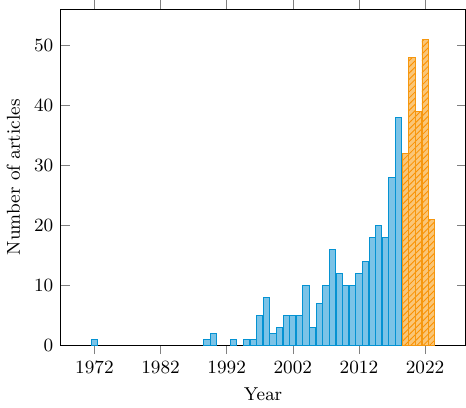}
    \caption{Distribution of contributions over years}%
    \label{fig:rq:timeline}
\end{figure}

The whole corpus of 457 primary studies we found was published in the years
1972--2023.
\Cref{fig:rq:timeline} displays the histogram of publications
per year\appendixtable{tab:last_five_years}
and highlights that AI applications on FM
appear to mirror the trend of general AI applications mentioned in \cref{sec:introduction}.
That is, we can observe a steadily growing interest in the field, especially since around 2005.

Focusing on the last five years only (recall \cref{sec:filter-years}), there still appears to be growing interest overall,
while not each year surpasses the respective previous one.
We have 32 contributions for 2019, 48 for 2020, 39 for 2021, 51 for 2022, and 21 for 2023.
Noteworthy are two observations:
First, for the last five years alone, there has been no strict upward trend in the number of publications,
We can observe a decline from 2020 to 2021.
Second, there were few publications in 2023. We explain this because we searched in the last quarter of 2023, when not all possible studies were published.
Given that we do not yet have complete data for 2023, it is impossible to forecast whether the overall trend will continue to grow or if the current peak from 2022 marks a global maximum.

\subsubsection{RQ~\ref{rq:pub-types}: \rqtext{rq:pub-types}}
\begin{figure}[!h]
\centering
    \includegraphics{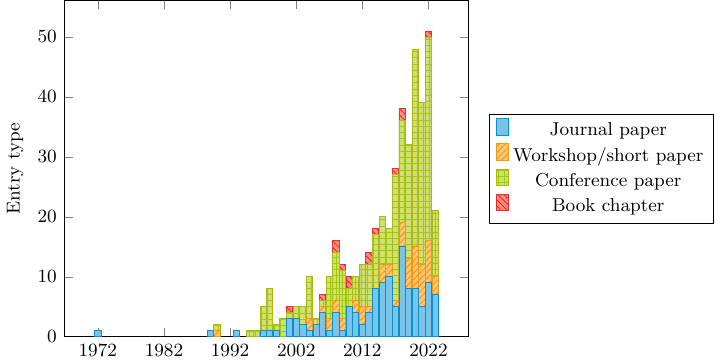}
    \caption{Distribution of publication venues by submission type}%
    \label{fig:rq:pub-types}
\end{figure}

\Cref{fig:rq:pub-types}\appendixtable{tab:publication_type}
displays the publications over the years, divided by publication venues.
The majority of studies were published as conference papers
while journal papers seem to be on the decline again since their
peak in 2018.
Within our five-year observation range,
124 contributions were conference papers, followed by 37 journal articles
29 workshop or short papers,
and only one (1) book chapter with no full books.

\begin{figure}[!h]
\centering
    \includegraphics{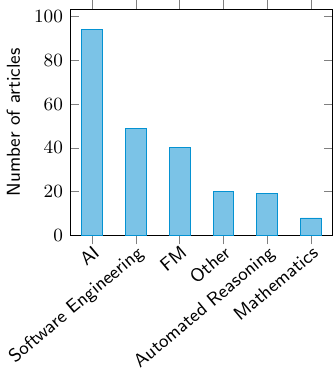}
    \caption{Distribution of publication venues by area}%
    \label{fig:rq:pub-types-area}
\end{figure}

\Cref{fig:rq:pub-types-area} shows the publication venues sorted by the main targeted area. For this, we skimmed through the titles of the conferences and sorted them into six different areas. We can see the majority of publications (94)
happened in AI-focused venues,
49 contributions were published in software engineering venues,
40 in FM venues,
19 in venues concerned with automated reasoning specifically,
8 in mathematics-focused venues,
and 20 in other, less topical venues.
Note that some venues have more than one area of focus,
e.g., we counted publications from the conference on artificial intelligence and theorem proving (AITP) for both AI and FM\@.

\subsubsection{RQ~\ref{rq:contributions}: \rqtext{rq:contributions}}%
\begin{figure}[ht]
    \centering
    \begin{subfigure}[T]{.5\textwidth}
        \centering
        \includegraphics{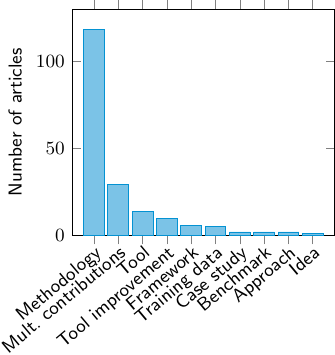}
        \caption{Singular contribution types}%
        \label{fig:rq:contributions}
    \end{subfigure}\hfill
    \begin{subfigure}[T]{.5\textwidth}
        \centering
        \includegraphics{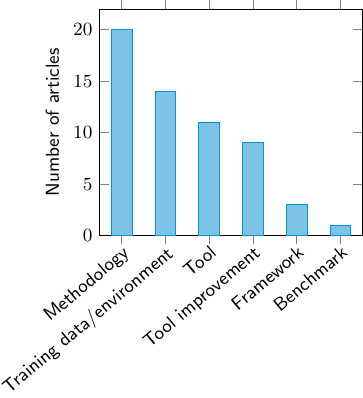}
        \caption{Division of studies' multiple contributions}%
        \label{fig:rq:contributions-multi}
    \end{subfigure}%
    \caption{Distribution of contribution types of collected primary studies.}
\end{figure}

\label{sec:rq:contributions}
Following the definitions of contribution types from
\cref{sec:contribution-classification}, we see the division of the studies into these contribution types in
\cref{fig:rq:contributions}\appendixtable{tab:contribution_type}.
We can see that the majority of contributions
were methodologies (118), i.e., practical application of AI to FM\@,
followed by studies with multiple types of contributions (29).
Third place was tool contributions (14), followed by improvements for tools (10), followed by frameworks (6),
training data (5), case studies (2), benchmarks (2), approaches (2), and finally, one (1) idea contribution.
Studies with multiple contribution types most commonly provide a
methodology (20/29),
training data (14/29),
or a tool (11/29).
See \cref{fig:rq:contributions-multi} for a complete breakdown.

\subsubsection{Intersecting \cref{rq:pub-types,rq:contributions}}
\begin{figure}[ht]
    \centering
    \includegraphics{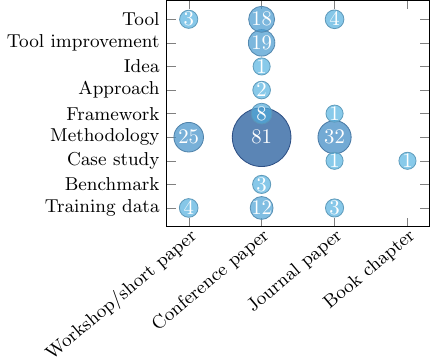}
    \caption[Contribution type and publication venues]{
        Contribution type and publication venues.
        Studies with multiple contribution types are considered individually for each distinct contribution type.
    }%
    \label{fig:type_venues}
\end{figure}

We can gain some additional insights by investigating the individual results in context with each other.
\Cref{fig:type_venues}\appendixtable{tab:contribution_publication_type} shows how the publication venues and the contribution types correlate.
We can see that for any contribution type, a heavy focus lies on conference papers.
Interestingly,
we can see
that case studies are only conducted within the scope of journal papers and book chapters.
Furthermore, workshops and short papers are primarily used to introduce new methodologies, which seem unexpected, as usually, due to limited space, authors aim to restrict themselves to outlining an approach or an idea.

\subsection{RQ~\ref{rq:quality}: \rqtext{rq:quality}}%
\label{subsec:quality}

\subsubsection{RQ~\ref{rq:ai-types}: \rqtext{rq:ai-types}}

\begin{figure}
    \centering
    \begin{subfigure}[T]{0.48\textwidth}
        \centering
        \includegraphics{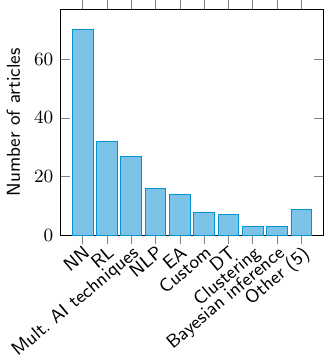}
        \caption{Quantity of the AI types in contributions}%
        \label{fig:ml_types}
    \end{subfigure}\hfill
    \begin{subfigure}[T]{0.48\textwidth}
        \centering
        \includegraphics{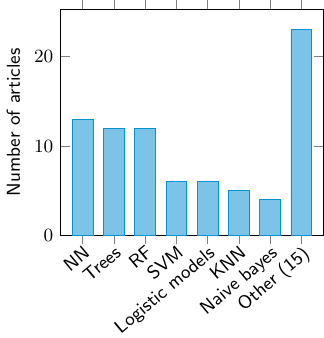}
        \caption{Detail view for the category \enquote{Mult.\ algorithms} of \cref{fig:ml_types}}%
        \label{fig:mul_ml_types}
    \end{subfigure}
    \caption[Overview of AI contributions ]{
        Overview of AI contributions.
        Underrepresented AI types are aggregated as \emph{Other}
        with a number in parentheses indicating how many sub-techniques
        were included.
    } %
    \label{fig:rq:2_1}
\end{figure}

In \cref{fig:ml_types}\appendixtable{tab:ai_types}, we can see that the majority of contributions use NN (70), followed by RL (32).
27 contributions use multiple techniques, while 16 use NLP methods and 14 use EA\@.
A total of 8 studies presented custom algorithms.
The rest of the contributions are divided between utilizing
decision trees (7),
clustering (3),
Bayesian inference (3),
KNN (2),
random forests (2),
data mining approaches (2),
automaton learning (2),
and SVM (1).

Studies that employed multiple algorithms mostly did so in a contrasting manner, i.e., they trained on numerous models to see which one performed best for their respective applications.
A detailed view of the category of multiple algorithms is given in \cref{fig:mul_ml_types}\appendixtable{tab:multp_ai_types}. Here, we can see that if multiple techniques are used, they mainly rely on NN (13/31), random forest (12/31), and some variant of tree learners (12/31).
Interestingly,
SVM (6/31) and KNN (5/31) applications are more predominant in a setting with multiple algorithms compared to being the sole focus of a study. We explain this by KNN or SVM, which are simple algorithms that authors can train quickly. Further, they need less fine-tuning than NNs, making them a baseline comparison approach.


\begin{figure}[h]
\centering
    \includegraphics{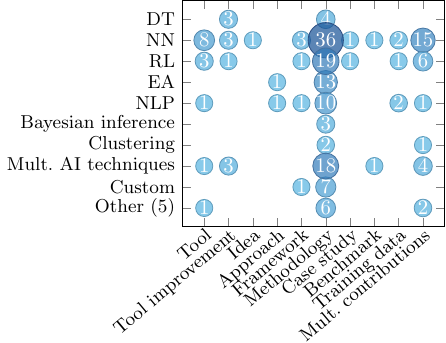}
    \caption{Distribution of contribution type and AI techniques}%
    \label{fig:ml_types_contributions}
\end{figure}

\paragraph*{Intersecting \cref{rq:ai-types,rq:contributions}}
\Cref{fig:ml_types_contributions}\appendixtable{tab:ai_vs_contributions} shows the intersection of \cref{rq:ai-types,rq:contributions}. The wide focus on methodologies was already assessed with \cref{fig:rq:contributions}. In \cref{fig:ml_types_contributions}, we can see that the amount of NN contributions remained focused on methodologies (37/189) or as part of a broader application of techniques (13/189) or a tool (8/189).

\begin{figure}[ht]
    \centering
 \includegraphics{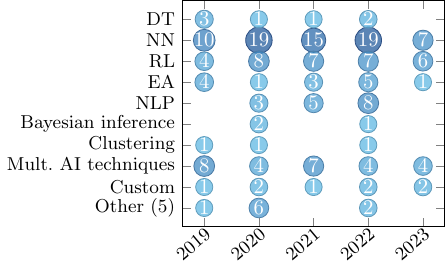}
    \caption{Distribution of year and AI techniques}%
    \label{fig:ml_types_years}
\end{figure}

\paragraph*{Intersecting \cref{rq:ai-types,rq:timeline}}
\Cref{fig:ml_types_years}\appendixtable{tab:ai_vs_years} show the intersection of \cref{rq:ai-types,rq:timeline}. Here, we can see that the amount of NN contributions remained steady over the years, similar to the amount of RL contributions. Noteworthy is that there is no notable decline in any technique over the observed period. The alternating nature of the overall amount of contributions was already found back in \cref{subsec:quantity}.

\subsubsection{RQ~\ref{rq:fm-types}: \rqtext{rq:fm-types}}
\begin{figure}
    \centering
    \begin{subfigure}[T]{0.48\textwidth}
        \centering
        \includegraphics{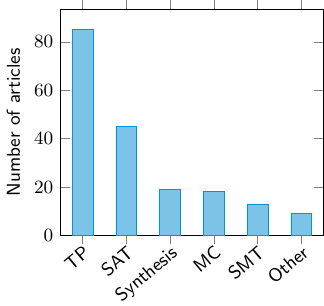}
        \caption{Quantity of targeted FM types}%
        \label{fig:fm_techniques}
    \end{subfigure}
    \begin{subfigure}[T]{0.48\textwidth}
        \centering
        \includegraphics{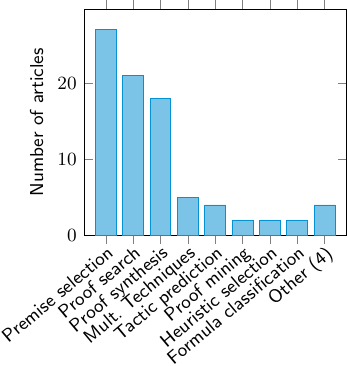}
        \caption{Quantity of targeted theorem-proving topics}%
        \label{fig:tp_techniques}
    \end{subfigure}\\
    \bigskip
    \begin{subfigure}[T]{1\textwidth}
        \centering
        \includegraphics{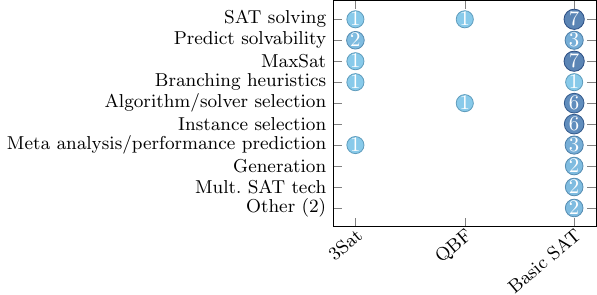}
        \caption{Quantity of targeted SAT topics}%
        \label{fig:sat_techniques}
    \end{subfigure}\\
    \bigskip
    \begin{subfigure}[T]{0.48\textwidth}
        \centering
        \includegraphics{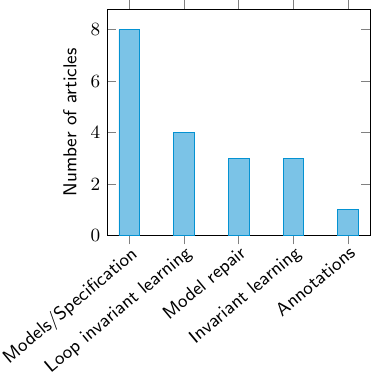}
        \caption{Quantity of targeted synthesis topics}%
        \label{fig:synth_techniques}
    \end{subfigure}%
    \label{fig:rq:2_2}
    \caption{Quantity of FM approaches with a detailed view of the most relevant contribution types}
\end{figure}

\Cref{fig:fm_techniques}\appendixtable{tab:fm_types} gives an overview of the FM techniques in the contributions. TP is leading by a large margin with 85 entries. SAT is next with 45 entries.
Synthesis approaches come in third place with 19 entries, followed by model checking (18) and SMT solving (13).
The remaining entries fall into algorithm selection, program analysis, and termination analysis.

In \cref{fig:tp_techniques}\appendixtable{tab:tp_types}, we can see a more detailed investigation of the topic of theorem proving. Here, we can see that premise, axiom, and clause selection received the most attention (27/85), followed by proof search (21/85) and proof synthesis (18/85). After that, a significant gap exists in a row of smaller topics.

In \cref{fig:sat_techniques}, we see the division of SAT approaches into their application areas. Overall, 37/45 studies focused on typical SAT solving, 2 on QBFs, and 6 on 3SAT\@.
Out of 45 articles,
9 focused on solving, 8 on finding MaxSAT, 7 on solver selection, 6 on instance selection, 5 aimed to predict solvability, and 4 contributions aimed at analyzing the performance of solvers. A small group of contributions aimed to do multiple things, generated SAT problems, and developed branching heuristics aimed at parameter selection or dependency analysis.

\Cref{fig:synth_techniques}\appendixtable{tab:synthesis_types} takes a closer look at the topic of synthesis. 8/19 contributions aimed to synthesize a model or specification, 4/19 aimed to learn a loop invariant, 3/19 aimed to repair models, and 3/19 targeted general invariant learning. One contribution targets the generation of annotations for the verification of JML code.

\begin{figure}
    \centering
    \begin{subfigure}[T]{0.48\textwidth}
        \includegraphics{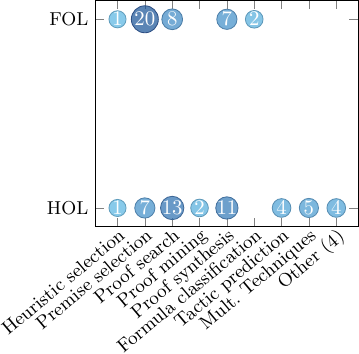}
        \caption[
            Division of TP applications into first-order logic
            and higher-order logic.
        ]{
            Division of TP applications into first-order logic (FOL)
            and higher-order logic (HOL).
        }%
        \label{fig:tp_logic_application}
    \end{subfigure}\hfill
    \begin{subfigure}[T]{0.48\textwidth}
        \includegraphics{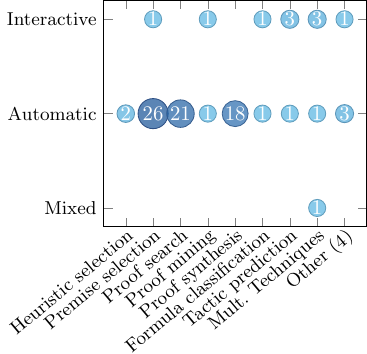}
        \caption{Automation grade against the type of TP subcategory}%
        \label{fig:tp_automatization}
    \end{subfigure}%
    \label{fig:detail_tp}
    \caption{Theorem proving in detail}
\end{figure}

\paragraph*{Detailed view on theorem proving}
A closer inspection is warranted, as TP makes up the most significant portion of contributions.
For the studies concerned with TP, we distinguished between higher-order logic (47/85) and first-order logic (38/85).
We further classified them by their automation levels:
fully automatic TP (74/85), interactive TP (10/85), and contributions employing both (1/85).
This granularity gives rise to additional observations about the problem structure.
\Cref{fig:tp_logic_application}\appendixtable{tab:tp_logic_app_types} shows that the majority of contributions (26/85) were made in the area of premise selection for automatic theorem proving, followed by proof search (21/85) and proof synthesis (18/85).

\Cref{fig:tp_automatization}\appendixtable{tab:tp_auto_app_types} shows that of the 38 first-order entries, there is a tight focus on premise selection (20/38), proof search (8/38), and proof synthesis (7/38). Although not as drastic, a similar spread can be observed for the higher-order contributions. Premise selection has 7/47 entries, proof search 13/47, and proof synthesis 11/47 entries. 4/47 entries focused on tactics prediction.

\begin{figure}
    \centering
    \begin{subfigure}[T]{0.48\textwidth}
        \includegraphics{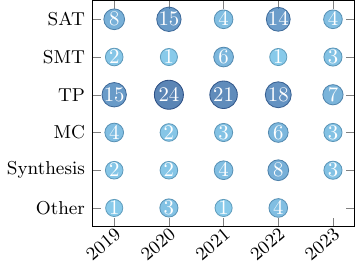}
        \caption{Distribution of FM techniques over the years }%
        \label{fig:fm_techniques_years}
    \end{subfigure}
    \begin{subfigure}[T]{0.48\textwidth}
        \includegraphics{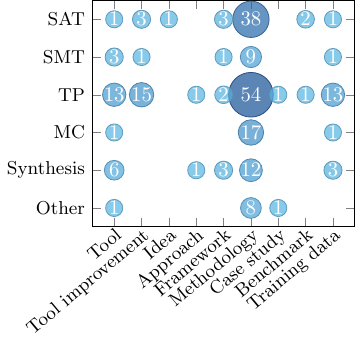}
        \caption[Distribution of FM techniques over contribution types]{
            Distribution of FM techniques over contribution types.
            Studies with multiple contribution types are considered
            for each distinct contribution type.
        }%
        \label{fig:fm_techniques_contribution_type}
    \end{subfigure}%
    \label{fig:rq:2_1_1_1_1_3}
    \caption{Quantity of FM approaches and comparison with time frame and contribution type}
\end{figure}

\paragraph*{Intersecting \cref{rq:fm-types} with \cref{rq:timeline,rq:contributions}}
\Cref{fig:fm_techniques_years}\appendixtable{tab:fm_years} shows the different types of FM over the years.
We can see no particular concentration on one year,
similar to \cref{fig:ml_types_years},
However, there may be some trends.
SAT had two solid years in 2020 and 2022, while TP had a slow decline; contributions to model checking and synthesis increased.
Nonetheless, given the small time frame of five years, any actual trends might be invisible to us.

In \cref{fig:fm_techniques_contribution_type}\appendixtable{tab:fm_contributions}, we showcase the distribution of FM types over contribution types.
While all FM techniques have found AI applications in the last five years,
we now see a fragmented distribution of specific contribution types per discipline.
In general, for TP, the different types of available contributions are the broadest, while for model checking, we are restricted to methodologies and one tool.
The notable absence of tool improvements outside of TP
might indicate developments in the respective subfields.
There might be no developments that deem an extension of tools necessary,
or no relevant tools can be extended.
However, the underlying reasons for this observation are out of the scope of this mapping study.
For the synthesis of formal models, we see a slightly better situation regarding a broader landscape of contribution types.
Overall, however, the absence of case studies and benchmarks
deserves attention and might highlight that the field is still in the early stages
of development.

\subsubsection{RQ~\ref{rq:ml-to-fm-dist}: \rqtext{rq:ml-to-fm-dist}}
\begin{figure}
    \centering
    \begin{subfigure}[T]{0.48\textwidth}
        \includegraphics{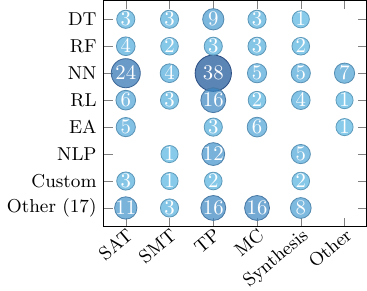}
        \caption{FM techniques with applied AI techniques}%
        \label{fig:fm_ml}
    \end{subfigure}\hfill
    \begin{subfigure}[T]{0.48\textwidth}
        \includegraphics{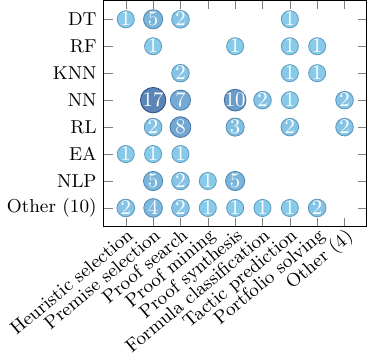}
        \caption{Theorem-proving techniques with applied AI techniques}%
        \label{fig:tp_ml}
    \end{subfigure}\\
    \bigskip
    \begin{subfigure}[T]{0.48\textwidth}
        \includegraphics{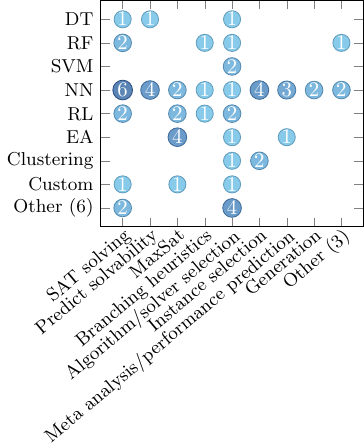}
        \caption{SAT with applied AI techniques}
        \label{fig:sat_ml}
    \end{subfigure}\hfill
    \begin{subfigure}[T]{0.52\textwidth}
        \includegraphics{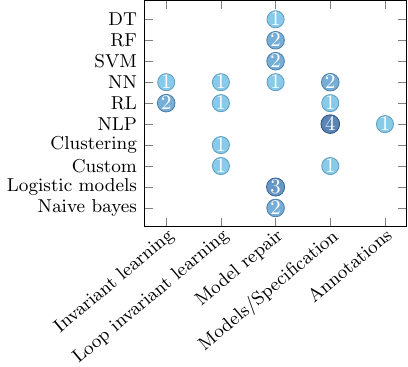}
        \caption{Synthesis with applied AI techniques}%
        \label{fig:synth_ml}
    \end{subfigure}%
    \caption[Quantity of FM approaches and comparison with time frame and contribution type]{
            The quantity of FM approaches and comparison with a time frame and contribution type. Studies that applied multiple AI techniques are considered individually for each distinct AI technique. Underrepresented techniques are aggregated as \emph{Other} with a number in parentheses indicating how many sub-techniques
            were included.
        }%
    \label{fig:rq:2_3}
\end{figure}

In \cref{fig:rq:2_3}, we see several overlays of FM applications with FM techniques at different granularities. \Cref{fig:fm_ml}\appendixtable{tab:fm_ai_contributions} is the most abstract.
Here, we can see that the bulk of the theorem-proving techniques apply
NNs (38), 
RL (16), 
and
NLP (12). 
SAT solvers mainly utilize
NNs (24).  
Model checking frequently uses
EA (6) and NN (5). 
The area of synthesis is mainly divided between
NNs (5), RL (4), and NLP (6)
as well.

Taking a closer look at TP in \cref{fig:tp_ml}
\appendixtable{tab:tp_ai_contributions}, we can see that NN is primarily used for the
premise, axiom, and clause selection (17) 
proof search (7) 
and proof synthesis (10). 
RL techniques are often used for
proof search as well (8). 
With three exceptions for heuristic selection, premise selection, or proof search,
EA is absent from the area of theorem proving.

Focusing on SAT in \cref{fig:sat_ml}%
\appendixtable{tab:sat_ai_contributions}, we can see that NN is dominant in most subcategories, except for portfolio selection and branching heuristics.
EA has higher relevance for MaxSat.

Finally, taking a closer look at the synthesis topic in \cref{fig:synth_ml}, we can see that NLP is relevant for the model generation
\appendixtable{tab:synth_ai_contributions}
and the generation of verification annotations.
NN and RL are mainly used for (loop) invariant learning.
For the repair of formal models, the contributions consider multiple techniques.

\subsubsection{RQ~\ref{rq:datasets}: \rqtext{rq:datasets}}
Lastly, we investigated the data sets that were found. Here, multiple observations took place.

\begin{figure}[htb]
    \centering
    \includegraphics{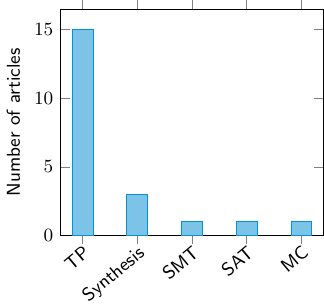}
    \caption{Overview of the distribution of data sets over FM domains}%
    \label{fig:fm_data}
\end{figure}

First, we searched for high-quality data sets enabling other researchers to conduct their research efficiently. As pointed out earlier, we found 21 data sets.
In \cref{fig:fm_data}\appendixtable{tab:datasets}, we see the respective scopes of the data sets.
Of the 21 data sets,
15 focused on TP, while 3 aimed at synthesis, 1 at SAT, 1 at SMT, and 1 at MC\@.

However, we were also interested in datasets that were not novel contributions but were found to be reused in other studies. For this, we skimmed all our studies for dataset mentions.
As it turns out,
of the 189 articles,
124 used external data sets.
The 65 remaining studies generate random samples or do not mention
their data source.
Noteworthy is that when random data is used, the generated data is seldom shared, hindering the reproduction of results.

The datasets found for the 124 contributions are largely heterogeneous. This means that most studies utilized different datasets.
However, we could find some data repositories that had multiple usages.
The Mizar Library\footnote{\url{https://mizar.uwb.edu.pl/library/}} (and associated libraries like, e.g., MPTP\cite{Urban2003}) saw 23 usages,
TPTP Problem Library\footnote{\url{https://tptp.org/TPTP/}} has 7 mentions,
The ProB machine library\footnote{\url{https://prob.hhu.de/w/index.php?title=Download}} got 5 mentions,
HoL\footnote{\url{https://isabelle.in.tum.de/library/}} 5,
SMTLib\footnote{\url{https://smt-lib.org}} 5,
SATComp\footnote{\url{https://satcompetition.github.io}} 5,
SATLib\footnote{\url{https://www.cs.ubc.ca/~hoos/SATLIB/benchm.html}} 4,
CoqGym\footnote{\url{https://github.com/princeton-vl/CoqGym}} 4.
The full list is available in the repository
linked in \cref{sec:introduction}.

\section{Discussion}%
\label{sec:discussion}

In the following,
we will evaluate the results and conclude by answering our research questions.
Further, we will outline lessons learned to provide actionable insights
for conducting future studies in the area.

\subsection{RQ~\ref{rq:demographics}: \rqtext{rq:demographics}}

In \cref{rq:timeline,rq:pub-types,rq:contributions}, we investigated the publication timeline for possible trends,
and assessed the qualitative nature of the studies,
i.e., the maturity of the publication format as well as
their level of postulated contribution,
following our taxonomies as defined in \cref{sec:contribution-classification}.
While we were able to analyze \cref{rq:timeline,rq:pub-types} over the full corpus of 457 publications, we restricted ourselves to the processed corpus of 189 publications
from the last five years
for content assessment as outlined in \cref{sec:filter-years}.
Consequently,
\cref{rq:contributions} is limited to these 189 publications.

For \cref{rq:timeline}, we see a general growth in the research field.
However, from the presented data points, it cannot be inferred alone
whether this trend follows the overall trend of AI's rise in popularity or is independent.

For \cref{rq:pub-types}, we see a strong focus on
conference papers, short papers, and workshops.
Together with the singular book chapter, this indicates an emerging field,
as there is a tendency to rely on publication methods supporting short review cycles.
This is further supported by the fact that the only book chapter was a case study.
Furthermore, we see that the AI side of things mainly drives the topic, and especially FM-based conferences publish less content about the application of AI in FM. This could indicate hesitance to adapt or a lack of AI maturity.
Contrasting this with \cref{rq:timeline}, we assume that the research interest will at least not plummet in the foreseeable future, but might plateau for the time being.

For \cref{rq:contributions}, we can see that the bulk of the work lies within the area of methodologies.
This could be due to the current paper requirements, which request a novel approach and some evaluation within the paper.
Another reason might be AI's applicative nature, which often allows it to serve as a support system to automate parts of the work. We also want to highlight the possibility that we primarily introduced methodology papers based on the nature of our search query, which focuses on the application of AI in the FM domain. However, due to rigorous snowballing, we are confident we correctly assessed the state of research in this field.

What is more interesting is what we do not see: many case studies and benchmarks. Recalling \cref{fig:type_venues}, we only see two publications providing case studies and two providing benchmarks. This may hint at a further gap in maturity, especially as these contribution types are primarily present in journal and book contributions.
One would assume more such publications if a set of robust solutions
for everyday problems were readily available so that researchers could compare their effectiveness against each other.
It also highlights the need for proper and uniform benchmark sets within the community, which can be shared between different research groups.

\subsection{RQ~\ref{rq:quality}: \rqtext{rq:quality}}

In \cref{rq:quality}, we were interested in how the AI application is distributed in FM\@.
For this, we posed \cref{rq:ai-types,rq:datasets,rq:fm-types,rq:ml-to-fm-dist} to investigate which AI techniques find appreciation in the community, which FM areas are already subject to AI application, if certain AI techniques are predominantly used in specific FM fields,
and whether there exist publicly available data sets that can be used in further research.

For \cref{rq:ai-types}, we saw in \cref{fig:rq:2_1} that most studies use NNs, followed by RL, and NLP and EAs are far behind.
From a paradigmatic point of view, it makes sense to rank these four at the top, as they all attempt to solve different problem settings.
NNs are a stand-in for classification and regression tasks,
RL is applied for learning behavioral policies,
NLP processes texts resembling natural language or code by extension, and EAs excel in optimization problems.
However, it is interesting to see NNs being so far ahead of the rest with
83 publications utilizing them.
Especially, since they are not the only algorithms suitable for classification and regression tasks, but competing algorithms such as KNN, SVM, or decision trees
are, at least as solely applied AI techniques, vastly underrepresented.
These algorithms, which we categorized as classical approaches in
\cref{sec:background-ai},
are, however, not forgotten, as we can see in
publications that contrast multiple different AI algorithms
(\cref{fig:mul_ml_types}).
This focus on NNs when only one technique is used can have multiple underlying
reasons.
First, implementing, training, and fine-tuning NNs is time-consuming and might require resources needed to investigate further algorithms.
Second, we have seen in \cref{fig:rq:timeline} that the research area gained traction in the early 2010s.
As this coincides with the advent of deep learning~\cite{sejnowski2018deep}, this might suggest the need for NNs to overcome the complexities the area of FM offers, and that more classical ML algorithms are unsuitable for most tasks. Third, the community might be biased towards using NNs from the start instead of applying more straightforward techniques first.

However, the prominence of more classical algorithms in the case of multiple AI techniques shows that the community seems to be aware that other algorithms might be
suitable for their problems as well.
This is especially amplified by \cref{fig:ml_types_contributions}, showing that most frequently, methodologies rely on various techniques to solve problems.

Regarding maturity,
\cref{fig:ml_types_years} indicates no noteworthy increase in techniques
besides NLP\@.
For 2023, we established insufficient data points so that we may see a rise in NLP-based contributions, which would fit a more general advent of NLP-related contributions due to the rise of transformer architectures and LLMs.

For \cref{rq:fm-types}, we have seen in \cref{fig:fm_techniques}
that TP and SAT are the predominant FM areas in which AI is applied,
making up 68.8\,\% of all publications in our five-year corpus.
We explain this by the fact that both areas are also of interest to non-FM
communities.
A detailed view of the authors' origin domains would be necessary to verify this assumption.
However, this is still beneficial for the FM community as a whole.
For instance,
\cref{fig:tp_logic_application}
shows a strong focus on FM-relevant daily use cases.
\Cref{fig:tp_automatization} reveals much current work regarding first-order logic on the especially relevant FM topics of premise selection.
While this trend is good, complex FM models rely on complicated data types and require sophisticated reasoning.

Regarding SAT, we can take away from \cref{fig:sat_techniques} that much focus lies directly on SAT solving. At the same time, other authors also attempt to predict satisfiability, i.e., whether a formula is SAT or UNSAT. The latter seems to contradict the rigorousness needed in FM, as the probabilistic nature of AI approaches lacks the certainty of the predicted satisfiability. This can be a product of cross-fertilization, as SAT is also relevant for non-FM communities. Here, a reasonable estimate about a formula's SAT or UNSAT may be more appreciated than in the FM community.

\Cref{fig:synth_techniques} shows that the interest in synthesis is low compared to other topics and primarily focuses on model generation.
Given the recent rise of generative AI approaches, we see room for promising research in this area.

A similar low interest seems to be in model checking. \Cref{fig:fm_techniques_contribution_type} shows a relatively low number of contributions in that area. The direct comparison with the TP field is interesting, as MC also contributes very little to tool and tool contributions, hinting at less application-focused research.
\Cref{fig:fm_techniques_years} shows a relatively consistent interest in recent years,
while \cref{fig:fm_techniques_contribution_type} emphasizes how little is done for current practitioners in terms of
tool (improvements), benchmarks, training data, and case studies.

For \cref{rq:ml-to-fm-dist}, we saw in \cref{fig:fm_ml} that NNs are the undeniable favorite in almost all FM areas investigated.
We already discussed why NNs might be more popular than other algorithms suitable for the same problem tasks above for \cref{rq:ai-types}.
However, this again suggests that the FM community mainly focuses on solving classification or regression problems rather than execution policies (RL) or direct processing of texts or code (NLP). The slight exception we see is in model checking, which focuses mainly on EAs, which interpret model checking as a sort of optimization problem, i.e., finding a path to a counter-example. From building the larger corpus of 457 studies, we know from seeing various titles that we missed many EA papers due to our 5-year restriction. Presumably, EA was more prominently used before deep learning took off.

An interesting observation is the substantial absence of any data mining-related
results, which might solve other problems still not covered by
NNs, RL, NLP, or EA\@.
In the 189 investigated studies, only three applied data mining techniques.
It is again worth noting that we only investigated the years 2019--2023.
Hence, data mining may have been more widespread before that,
much like EA\@. More thorough data mining would be needed for a deeper insight, and only a secondary study would be required.

Finally, for \cref{rq:datasets}, we found 19 datasets (or training environment) contributions in our corpus of 189 studies.
Considering the two separate, data set-only publications that did not pass \cref{ic:1} (see  \cref{sec:filter-years}), this leaves us with 21 total data set contributions.

With the analyzed high heterogeneity of the used datasets, as already mentioned in \cref{sec:rq:contributions}, there may be a general underlying issue. There seems to be no defined benchmark \textit{gold standard} against which to run new insights. Instead, results are run on the data sets at hand. Furthermore, while datasets are publicly available for TP and SAT solving, the number of usable, established, and available benchmark datasets for other research areas that fit the named criteria is minuscule.

Generally, we see a current trend where researchers use datasets as they see fit or even generate data without referencing the procedure or providing the dataset afterward.
This trend fits the overall trend of AI being only weakly reproducible~\cite{hutson2018reproducibility}. However, this trend endangers the whole research branch, as reproducibility is a core part of science.

In conclusion, we strongly need unified defaults to allow reproducibility and introduce familiar benchmarks into the field.
Possible issues hindering such a development might be found in the various
existing formalisms, which are not necessarily used or even supported
by individual research groups.
Such boundaries, of course, weaken the impact a data set publication might have
onto researchers working with differing formalisms.

\subsection{Observed objectives of AI applications in FM}
Generally, we see that the stark majority of publications use AI as an assistance tool to enable or enhance a respective FM tool's performance. That is, AI is used to find models, clauses, or lemmas. However, the found artifacts are then still checked by a formal tool.

With the notable exception of SAT solving,
where 5 contributions directly predict the satisfiability of a formula with AI, we did not observe any focus on finding AI-generated guarantees for formal models to replace an FM tool.
This seems only natural, as FM are about believable and reproducible guarantees. Therefore, giving most AI approaches a more prominent role comes with the price of admission: introducing determinism and uncertainty. For strictly formal-safety-related projects, this may be too much of a burden to pass.

Another observation we did not make was the presence of auto-active verification. Analyzing existing program code did not draw much attention, and the only remotely connected contribution was outside our search area.
\subsection{Suggested research directions}


From the given analysis, we can derive five areas of growth potential,
which we outline below.

\paragraph*{Development of unified benchmarks and data sets}
As pointed out in our discussion to \cref{rq:datasets}, we are concerned with the lack of established data sets and benchmarks in the FM sub-communities. While many contributions have already tackled this problem for TP, the other areas seem to fall short. This creates room for reproducibility issues between research groups and hinders a more precise comparison between results and established benchmarks that could be provided.

Developing strong baselines might increase comparability between different methods and reduce the entry hurdle for new researchers due to publicly available data sets. The main challenge is to create data sets that can be used for different formalisms. This would allow smaller communities to still benefit from the available data, which might be presented in a more popular formalism while furthering research in their area.

An orientation on how such a unified benchmark environment could look would be OpenAIGym~\cite{Brockman2016}, a Toolkit for RL to foster research by enhancing reproducibility.

\paragraph*{Investigate the benefits of data mining for FM}
In the age of big data, we were expecting more studies that apply data mining techniques,
be it for feature engineering or performance analytical approaches. This was not the case.
While it might very well be that the kind of problems solvable by data mining are irrelevant to the FM communities, it is also possible that potential benefits are not well-known.

We propose to research whether and how data mining approaches can benefit FM.
First, this would highlight potential benefits to the community. Second, due to the investigative nature of data mining,
this might strengthen our insights into our tools and best practices.
Consequently, this might further the quality of other AI applications in FM due to new knowledge.

\paragraph*{Application of LLMs and generative AI}
While LLMs, at this point, are established tools that have found their way into daily applications, we have found little use in our corpus of studies.
This can be due to their relatively high employment costs.
Another reason could be that respective studies are still published and invisible to our systematic mapping study.
Nevertheless, we see strong potential for generative AI approaches,
especially in synthesis and auto-formalization. However, simple test cases or documentation generation for existing models seems a promising starting point for building trust.

We envision substantial benefits from applying generative AI during FM development and highlight the need for more research.
One can easily envision
tool chains of automatic generation, checking, and documentation of formal models, logical formulas, or proofs with reproducible and provable results.
However, we also see the need for more exploration regarding how generative AI can be
helpful besides text production.
For instance,
how could generative AI be leveraged in proof mining?
What stages in the formal development process can benefit from generated inputs, and what types could and should be generated? Generative AI for FM may be thought of further than pure text and code generation and more toward reasoning support.

\paragraph*{Increase the potential of AI in model checking}
Another underrepresented topic is model checking. Anecdotally, we know model checking had much EA applied in the past, while it was a seemingly unpopular application area for AI in our five-year corpus. However, we see significant potential in revisiting the research with a modern lens.
Especially in the context of bounded model checking (BMC)
or statistical model checking (SMC) and faulty state prediction.
BMC cuts off parts of a model's state space for performance and computability.
SMC checks states until a certain confidence in the absence of faulty states is reached. Both approaches do not fully traverse the state space and
can only indicate the absence of faults.

AI-enforced model checking could complement BMC, SMC, and model checking by predicting potential faulty states or needed search depths, learning search policies via RL to reach faulty states more targeted, or predicting equivalence classes of states for search space reduction. For example, after estimating a defective state or a search depth, the model checker could be applied in a very targeted manner to confirm the predicted state.

\section{Threats to validity}%
\label{sec:threats}
A study of this type and extent can only be conducted with the possibility of encountering threats that threaten validity.
\citet{Zhou2020} serves as a baseline for this discussion, as their contribution meticulously lists the possible threats to validity and how they may be addressed.

\paragraph*{Internal validity}
A potential threat to the conclusion could be our selection of articles
and our data source. To ensure the quality of our data, we carefully drafted our IC/EC, and in cases where we found them not yet sufficient, we enhanced them. Two experienced researchers, i.e., the first two authors,  carefully read the publications.

Our data source was four famous known databases containing the relevant literature~\cite{Dyba2007}.
Due to the extensive snowballing, we made sure that the potential for blind spots, i.e., publications that are only listed by one engine or none at all, is minimized. This also holds for the case of WoS, where different result sizes are obtained. The divergence between the two results was marginal, so we are confident that excessive snowballing alleviates it. The snowballing also alleviated some of our rigor by applying IC/EC
early in our search process. While we reduced our initial tally of 1492 publications to only 89, a reduction of two orders of magnitude,
snowballing helped us recapture significant contributions into our database again.

\paragraph*{Construct validity}
Another threat is the selection and use of search terms. We used PICO as \citet{Kitchenham2007} suggest to select the appropriate search term.
Nevertheless, we know it is impossible to cover all types of FM, AI, and all their acronyms in one query due to the technical limitations of the search engines and the practical feasibility of thoroughly skimming the amount of papers found this way. Therefore, the aim was to keep the initial set small and conduct an exhaustive snowballing procedure to include all relevant publications.

The amount of snowballing results was unsurprisingly large,
as snowballing does not suffer from a compromise for a search query.
This is a limitation already pointed out by \citet{Wohlin2014}.

\paragraph*{Conclusion validity}
While we followed \citet{Kitchenham2007} guidelines for a successful mapping study to minimize the threat to the conclusion of this study, one remaining potential threat is whether or not to include grey literature, as we could have missed crucial contributions that change the outcome of this study.
Grey literature was not considered for two main reasons.
First, we did not encounter large quantities of grey literature during our snowballing.
Thus,
We are positive that we can represent current research directions without considering them.
Second, we only did a soft skim of the content,
and therefore, we can not offer a quality and feasibility assessment of the included contributions.
As we aim to provide the reader with only high-quality studies,
we decided to require official publications.

\section{Related work}%
\label{sec:related_work}
While the lack of available literature partially motivated this work, there are noteworthy related works. One of the first contributions one finds when investigating the topic is that of \citet{Amrani2018}. Here, the authors came to a similar conclusion as we did about the complicated nature of the investigated landscape of available literature. Therefore, they also resorted to a complex search string. The evaluation and focus, however, are different. The authors limit themselves to an excerpt of initial results and only conduct backward snowballing, leaving room for whether the corpus is complete. Additionally, the research questions are more specifically tailored to individual FM topics, while our work aims to give a general overview.

\citet{Wang2020b} aims to give a general taxonomy of learning-based FM\@. One core contribution is that they divide the topic into two general directions: the learning of formal specifications and learning for formal verification. They proceeded to structure the literature found by individuals under this taxonomy. Compared to their work, our work aims to give a general introduction to the cross-section of the two fields, while the authors dive into the specifics of learning algorithms for formal verification. 

\paragraph*{Solvers and solving}
Large quantities of work exist to evaluate the use of AI to enable solvers. A general overview was given by \citet{Ganesh2023}. In the context of SAT, multiple works exist. A general overview with comments on the use of AI was presented by \citet{Kilani2013}. Explicit SAT solving was subject to two studies~\cite{Holden2021,Guo2023}. Configuration and benchmarking of SAT solvers were investigated in three contributions~\cite{Granmo2010,Hoos2021,Fuchs2023}.

We found many contributions that targeted the general constraint satisfaction problem during our search. However, we excluded these results as they are only loosely connected with FM@. \citet{Amadini2013} conducted a general evaluation of portfolio approaches for CSP problems. \citet{Popescu2022} gives an overview of machine learning techniques for CSP problems. 

\paragraph*{Proving}
Multiple works aim to provide an overview of the topic within the math and automated reasoning community. \citet{Urban2013} give an overview of AI-supported automated theorem proving. \citet{Blanchette2014} surveyed ML-supported axiom selection, and two years later, \citet{Blanchette2016} surveyed the rising interest in automatizing reasoning over proofs. \citet{England2018} published a survey on ML for symbolic reasoning. \citet{Tran2022} gave a short overview of the emerging field of NLP for premise selection.

\paragraph*{Program analysis}
For general program analysis, there are works for runtime prediction by \citet{Hutter2014}. \citet{Kumazawa2020} published a survey on applying swarm intelligence (a flavor of EAs) for software verification. \citet{Bennaceur2018} published on which ML techniques may be suitable for software analysis.

\paragraph*{Model/specification generation}
\citet{Brunello2019} and \citet{Buzhinsky2019} published surveys on creating temporal logic from natural language, while \citet{Fuggitti2023} and \citet{Szegedy2020} discuss the topic of auto-formalization of natural language requirements.

\paragraph*{Others} Less prominent were the topics of model repair, where \citet{Barriga2022} published a longer article about the state of the art. \citet{Haltermann2022} gives an overview of invariant generation. \citet{Pan2022} surveyed hardware vulnerability analysis via ML, a field to which we found no contributions. For model checking, \citet{Besbas2022} gave a brief literature review of four pages.

\section{Conclusion}%
\label{sec:conclusion}
In this systematic mapping study, we present the results of our work on assessing the quantity of research in applying artificial intelligence to formal methods. Overall, we found 457, from which we selected 189 for a closer investigation. Concluding from this investigation, we see the current trend is yet to mature, as many contributions are making some practical applications. However, only a few studies aim to create theoretical groundwork, benchmarks, or case studies. Furthermore, we see a focus on theorem proving and SAT solving. Model checking and model synthesis are underrepresented compared to this. Most work uses neural nets, reinforcement learning, or a combination of multiple approaches. Noteworthy is the predominant degradation of AI to a support function within existing formal methods.

\backmatter

\bmhead{Supplementary information}

The supplementary material provides an overview of the sorted, investigated primary studies. It consists of tables, the basis for the figures presented in this work.

%
%

\clearpage
\begin{appendices}

\section{Investigated studies}%
\begin{tabularx}{\textwidth}{@{}llXr@{}}
  \caption{Investigated primary studies sorted by their FM application}%
  \label{tab:sorted_primary_studies} \\
  \toprule
  FM technique & Application & Articles & Count \\
  \midrule
  \endfirsthead

  \caption{(Continued from previous page.)} \\
  \toprule
  FM technique & Application & Articles & Count \\
  \midrule
  \endhead

  \bottomrule
  \multicolumn{4}{r@{}}{\small (Continued on next page.)} \\
  \endfoot

  \midrule
  {Unique total} & & & 189 \\ 
  \bottomrule
  \endlastfoot

    SAT & SAT solving & \cite{Amizadeh2019a,Atkari2020,Fournier2022,Li2022,Nejati2020,Selsam2019a,Xu2022a,Yolcu2019,Zhang2022,sun2020neurogift} & 10 \\
     & Predict solvability & \cite{Cameron2020,Chang2022,Selsam2019,Xu2021,Zhang2020} & 5 \\
     & MaxSat & \cite{Berden2022,Chang2020,Framil2022,Kumar2023,Lassouaoui2019,Marino2021,Nurcahyadi2022,Sadeg2021,Zheng2022} & 9 \\
     & SAT parameter selection & \cite{Beskyd2022} & 1 \\
     & Branching heuristics & \cite{Kurin2020,Lorenz2020} & 2 \\
     & Algorithm/solver selection & \cite{Fournier2022,Fuchs2023,Lederman2020,Nejati2020,Richter2020,Sadeg2021,Sadreddin2022,Wang2019a} & 8 \\
     & Instance selection & \cite{Dalla2021,Danisovszky2020a,Farooque2023,Han2020,Hireche2019,Hireche2020} & 6 \\
     & Dependency analysis & \cite{Yan2023} & 1 \\
\end{tabularx}

\begin{tabularx}{\textwidth}{@{}lXr@{}}
  \caption{Datasets }%
  \label{tab:datasets} \\
  \toprule
  FM & Articles & Count \\
  \midrule
  \endfirsthead

  \caption{(Continued from previous page.)} \\
  \toprule
  FM & Articles & Count \\
  \midrule
  \endhead

  \bottomrule
  \multicolumn{3}{r@{}}{\small (Continued on next page.)} \\
  \endfoot

  \midrule
  {Total} & & 21 \\ 
  \bottomrule
  \endlastfoot

    SAT & \cite{You2019} & 1 \\
    SMT & \cite{Dunkelau2019} & 1 \\
    TP & \cite{Bansal2019,Blaauwbroek2020,Goertzel2022,Han2022,Jiang2021,Nagashima2020,Piotrowski2020a,PurgaL2021,Reichel2023,Shminke2022,Urban2020,Wang2020a,Wu2020,Wu2020a,Yang2019} & 15 \\
    MC & \cite{Wang2021b} & 1 \\
    Synthesis & \cite{Bordg2022,Cai2019,He2022} & 3 \\
\end{tabularx}

\ifshowappendix
\section{Tables for RQ~1}
\begin{tabularx}{\textwidth}{@{}lXr@{}}
  \caption{Entries grouped by years}%
  \label{tab:last_five_years} \\
  \toprule
  Year & Articles & Count \\
  \midrule
  \endfirsthead

  \caption{(Continued from previous page.)} \\
  \toprule
  Year & Articles & Count \\
  \midrule
  \endhead

  \bottomrule
  \multicolumn{3}{r@{}}{\small (Continued on next page.)} \\
  \endfoot

  \midrule
  {Total} & & 191 \\ 
  \bottomrule
  \endlastfoot

    2019 & \cite{Amizadeh2019a,Bansal2019,Blanchette2019,Cai2019,Cai2019a,Chen2019,Chvalovsky2019a,Dunkelau2019,Glorot2019,Goertzel2019,Goertzel2019a,Hireche2019,Holden2019,Huang2019,Jakubuv2019,Kumazawa2019,Lassouaoui2019,Ma2019,Nagashima2019,Nawaz2019,Nikolic2019,Rafe2019,Rawson2019,Selsam2019,Selsam2019a,Wang2019a,Yang2019,Yolcu2019,You2019,Zhang2019,Zhu2019,Zombori2019} & 32 \\
    2020 & \cite{Atkari2020,Bartek2020,Blaauwbroek2020,Brown2020a,Cameron2020,Cao2020,Chang2020,Danisovszky2020a,Dunkelau2020,Faerber2020,Ferreira2020,First2020,Fu2020,Gauthier2020,Gauthier2020a,Goertzel2020,Han2020,Hireche2020,Jaeger2020,Jakubuv2020,Kurin2020,Lederman2020,LeventiPeetz2020,Liu2020b,Lorenz2020,Mo2020,Nagashima2020,Nawaz2020,Nejati2020,Nie2020,Olsak2020,Paliwal2020,Piotrowski2020,Piotrowski2020a,Richter2020,Richter2020a,SanchezStern2020,Shigyo2020,Si2020,Urban2020,Wang2020,Wang2020a,Wu2020,Wu2020a,Yao2020,Zhang2020,Zombori2020,sun2020neurogift} & 48 \\
    2021 & \cite{Baghdasaryan2021,Bartek2021,Chvalovsky2021,Crouse2021,Dalla2021,Dunkelau2021,Firoiu2021,Goertzel2021,Han2021,Hula2021,Jiang2021,Kobayashi2021,Koscinski2021,Kumazawa2021,Laurent2021,Marino2021,Nawaz2021,Nawaz2021b,Piepenbrock2021,Pimpalkhare2021,Pimpalkhare2021a,Poesia2021,Puccetti2021,PurgaL2021,Qian2021,Rawson2021,Sadeg2021,Scott2021,Scott2021a,Suda2021,Suda2021a,Wang2021,Wang2021b,Wu2021,Xu2021,Yao2021,Zhang2021,Zombori2021,Zombori2021a} & 39 \\
    2022 & \cite{Abate2022,Allamanis2022,Ayguen2022,Berden2022,Beskyd2022,Bordg2022,Cai2022,Cao2022,Chang2022,Cherukuri2022,First2022,Fournier2022,Framil2022,Garzon2022,Giacobbe2022,Goertzel2022,Gross2022,Han2022,He2022,Hu2022,Huang2022,Janota2022,Jiang2022,Kumazawa2022,Lample2022,Laurent2022,Li2022,Liu2022,Luo2022,Mangla2022,Morris2022,Nayak2022,Nurcahyadi2022,Ozolins2022,Palermo2022,Piepenbrock2022,Pira2022,Sadreddin2022,Shminke2022,TsutomuKumazawa2022,Tworkowski2022,Wei2022,Welleck2022,Wu2022,Wu2022a,Xu2022a,Xu2022b,Zhang2022,Zheng2022,Zhu2022,Zhu2022a} & 51 \\
\end{tabularx}

\begin{tabularx}{\textwidth}{@{}lXr@{}}
  \caption{Entries grouped by type of publication venue}%
  \label{tab:publication_type} \\
  \toprule
  Entry type & Articles & Count \\
  \midrule
  \endfirsthead

  \caption{(Continued from previous page.)} \\
  \toprule
  Entry type & Articles & Count \\
  \midrule
  \endhead

  \bottomrule
  \multicolumn{3}{r@{}}{\small (Continued on next page.)} \\
  \endfoot

  \midrule
  {Total} & & 191 \\ 
  \bottomrule
  \endlastfoot

    Workshop/short paper & \cite{Bartek2020,Berden2022,Blanchette2019,Bordg2022,Brown2020a,Dunkelau2021,Firoiu2021,Glorot2019,Goertzel2019a,Han2020,Han2022,Holden2019,Hu2023,Jakubuuv2023,Koscinski2021,Laurent2021,Liu2020b,Morris2022,Palermo2022,Piepenbrock2021,Puccetti2021,Qian2021,SanchezStern2020,Tworkowski2022,Wu2020,Wu2020a,Wu2022a,Yan2023,Zombori2019} & 29 \\
    Conference paper & \cite{Abate2022,Amizadeh2019a,Atkari2020,Ayguen2022,Bansal2019,Bartek2021,Beskyd2022,Blaauwbroek2020,Cai2019a,Cameron2020,Cao2022,Chang2020,Cherukuri2022,Chvalovsky2019a,Chvalovsky2021,Chvalovskỳ2023,Crouse2021,Dalla2021,Danisovszky2020a,Dunkelau2019,Dunkelau2020,Farooque2023,Ferreira2020,First2022,Fournier2022,Framil2022,Garzon2022,Gauthier2020a,Giacobbe2022,Goertzel2019,Goertzel2020,Goertzel2021,Goertzel2022,Gross2022,Han2021,He2022,Huang2019,Huang2022,Hula2021,Jaeger2020,Jakubuv2019,Jakubuv2020,Janota2022,Jiang2021,Jiang2022,Kobayashi2021,Kumazawa2021,Kumazawa2022,Kumazawa2023,Kurin2020,Lample2022,Laurent2022,Lederman2020,Leeson2023,LeventiPeetz2020,Li2022,Liu2023,Lorenz2020,Luo2022,Luo2023,Ma2019,Mangla2022,Mo2020,Nagashima2019,Nagashima2020,Nawaz2019,Nayak2022,Nejati2020,Nie2020,Olsak2020,Ozolins2022,Paliwal2020,Piepenbrock2022,Pimpalkhare2021,Pimpalkhare2021a,Piotrowski2020,Piotrowski2020a,Poesia2021,Rawson2019,Rawson2021,Reichel2023,Richter2020a,Sadeg2021,Sadreddin2022,Scott2021,Scott2021a,Selsam2019,Selsam2019a,Shigyo2020,Si2020,Suda2021,Suda2021a,Urban2020,Wang2019a,Wang2020,Wang2020a,Wang2021,Wang2021b,Wang2023,Wang2023a,Wei2022,Welleck2022,Wu2021,Wu2022,Xu2021,Xu2022a,Xu2022b,Yang2019,Yao2020,Yao2021,Yolcu2019,You2019,Yu2023,Zhang2019,Zhang2020,Zhang2021,Zhang2022,Zhang2023,Zheng2022,Zhu2022,Zombori2020,Zombori2021,Zombori2021a,sun2020neurogift} & 124 \\
\end{tabularx}

\begin{tabularx}{\textwidth}{@{}lr@{}}
  \caption{Overview of venues' respective research areas}%
  \label{tab:venues_all} \\
  \toprule
  Venue focus & Submissions \\
  \midrule
  \endfirsthead

  \caption{(Continued from previous page.)} \\
  \toprule
  Venue area & Submissions \\
  \midrule
  \endhead

  \bottomrule
  \multicolumn{2}{r@{}}{\small (Continued on next page.)} \\
  \endfoot

  \midrule
  {Unique total} & 191 \\ 
  \bottomrule
  \endlastfoot

    AI & 94 \\
    Software Engineering & 49 \\
    FM & 40 \\
    Other & 20 \\
    Automated Reasoning & 19 \\
    Mathematics & 8 \\
\end{tabularx}

\begin{tabularx}{\textwidth}{@{}lXr@{}}
  \caption{Entries grouped by type of contribution}%
  \label{tab:contribution_type} \\
  \toprule
  Contribution type & Articles & Count \\
  \midrule
  \endfirsthead

  \caption{(Continued from previous page.)} \\
  \toprule
  Contribution type & Articles & Count \\
  \midrule
  \endhead

  \bottomrule
  \multicolumn{3}{r@{}}{\small (Continued on next page.)} \\
  \endfoot

  \midrule
  {Total} & & 189 \\ 
  \bottomrule
  \endlastfoot

    Tool & \cite{Cherukuri2022,First2020,Gauthier2020,Gross2022,Huang2019,Laurent2021,Leeson2023,Nie2020,Rawson2019,Rawson2021,SanchezStern2020,Scott2021a,Zombori2020,Zombori2021a} & 14 \\
    Tool improvement & \cite{Ayguen2022,Chvalovsky2021,Chvalovskỳ2023,Goertzel2019,Goertzel2020,Goertzel2021,Jakubuv2019,Piepenbrock2022,Suda2021a,Wang2019a} & 10 \\
    Idea & \cite{Xu2022a} & 1 \\
    Approach & \cite{Nagashima2019,Shigyo2020} & 2 \\
    Framework & \cite{Abate2022,Chang2022,Jiang2022,Kobayashi2021,Wu2021,Zhang2022} & 6 \\
    Methodology & \cite{Abdelaziz2023,Amizadeh2019a,Baghdasaryan2021,Bartek2020,Bartek2021,Berden2022,Beskyd2022,Blanchette2019,Brown2020a,Cai2019a,Cai2022,Cameron2020,Cao2020,Cao2022,Chvalovsky2019a,Crouse2021,Dalla2021,Danisovszky2020a,Dunkelau2020,Dunkelau2021,Faerber2020,Farooque2023,Ferreira2020,Firoiu2021,Fournier2022,Framil2022,Fu2020,Fuchs2023,Garzon2022,Gauthier2020a,Giacobbe2022,Glorot2019,Goertzel2019a,Han2020,Han2021,Hireche2019,Hireche2020,Holden2019,Hu2022,Hu2023,Huang2022,Hula2021,Jaeger2020,Jakubuuv2023,Janota2022,Koscinski2021,Kumar2023,Kumazawa2019,Kumazawa2021,Kumazawa2022,Kumazawa2023,Kurin2020,Lample2022,Lassouaoui2019,Laurent2022,Lederman2020,LeventiPeetz2020,Liu2020b,Liu2022,Lorenz2020,Luo2022,Luo2023,Ma2019,Mangla2022,Marino2021,Morris2022,Nawaz2019,Nawaz2020,Nawaz2021,Nawaz2021b,Nayak2022,Nejati2020,Nikolic2019,Nurcahyadi2022,Olsak2020,Ozolins2022,Palermo2022,Paliwal2020,Piepenbrock2021,Pimpalkhare2021a,Piotrowski2020,Pira2022,Poesia2021,Puccetti2021,Qian2021,Rafe2019,Richter2020,Richter2020a,Sadeg2021,Sadreddin2022,SanchezStern2023,Scott2021,Scott2023,Si2020,TsutomuKumazawa2022,Tworkowski2022,Wang2020,Wang2021,Wang2023,Wang2023a,Wei2022,Welleck2022,Wu2022,Wu2022a,Xu2021,Yan2023,Yao2020,Yao2021,Yolcu2019,Zhang2019,Zhang2020,Zhang2023,Zheng2022,Zhu2019,Zhu2022,Zhu2022a,Zombori2019,Zombori2021} & 118 \\
    Case study & \cite{Allamanis2022,Poesia2023} & 2 \\
    Benchmark & \cite{Atkari2020,Chang2020} & 2 \\
\end{tabularx}

\begin{tabularx}{\textwidth}{@{}llXr@{}}
  \caption{Entries grouped by type of contribution and publication venue}%
  \label{tab:contribution_publication_type} \\
  \toprule
  Contribution type & Entry type & Articles & Count \\
  \midrule
  \endfirsthead

  \caption{(Continued from previous page.)} \\
  \toprule
  Contribution type & Entry type & Articles & Count \\
  \midrule
  \endhead

  \bottomrule
  \multicolumn{4}{r@{}}{\small (Continued on next page.)} \\
  \endfoot

  \midrule
  {Total} & & & 189 \\ 
  \bottomrule
  \endlastfoot

    Tool & Workshop/short paper & \cite{Laurent2021,SanchezStern2020} & 2 \\
     & Conference paper & \cite{Cherukuri2022,Gross2022,Huang2019,Leeson2023,Nie2020,Rawson2019,Rawson2021,Scott2021a,Zombori2020,Zombori2021a} & 10 \\
     & Journal paper & \cite{First2020,Gauthier2020} & 2 \\
    Tool improvement & Conference paper & \cite{Ayguen2022,Chvalovsky2021,Chvalovskỳ2023,Goertzel2019,Goertzel2020,Goertzel2021,Jakubuv2019,Piepenbrock2022,Suda2021a,Wang2019a} & 10 \\
    Idea & Conference paper & \cite{Xu2022a} & 1 \\
    Approach & Conference paper & \cite{Nagashima2019,Shigyo2020} & 2 \\
    Framework & Conference paper & \cite{Abate2022,Jiang2022,Kobayashi2021,Wu2021,Zhang2022} & 5 \\
     & Journal paper & \cite{Chang2022} & 1 \\
\end{tabularx}

\section{Tables for RQ~2}

\begin{tabularx}{\textwidth}{@{}lXr@{}}
  \caption{Entries grouped by type of AI type used}%
  \label{tab:ai_types} \\
  \toprule
  AI technique & Articles & Count \\
  \midrule
  \endfirsthead

  \caption{(Continued from previous page.)} \\
  \toprule
  AI technique & Articles & Count \\
  \midrule
  \endhead

  \bottomrule
  \multicolumn{3}{r@{}}{\small (Continued on next page.)} \\
  \endfoot

  \midrule
  {Total} & & 189 \\ 
  \bottomrule
  \endlastfoot

    DT & \cite{Blanchette2019,Goertzel2019,Goertzel2020,Jakubuv2019,Janota2022,Lample2022,Xu2021} & 7 \\
    RF & \cite{Beskyd2022,Nejati2020} & 2 \\
    SVM & \cite{Richter2020} & 1 \\
    KNN & \cite{Blaauwbroek2020,Gauthier2020} & 2 \\
    NN & \cite{Allamanis2022,Baghdasaryan2021,Bansal2019,Bartek2021,Cameron2020,Cao2020,Cao2022,Chang2020,Chang2022,Cherukuri2022,Chvalovsky2021,Chvalovskỳ2023,Dalla2021,Danisovszky2020a,Farooque2023,Ferreira2020,First2022,Garzon2022,Giacobbe2022,Glorot2019,Goertzel2019a,Goertzel2022,Han2020,Han2022,He2022,Hu2023,Huang2019,Huang2022,Hula2021,Jakubuv2020,Kobayashi2021,Kommrusch2023,Laurent2021,Leeson2023,LeventiPeetz2020,Li2022,Liu2022,Luo2022,Marino2021,Nie2020,Olsak2020,Ozolins2022,Paliwal2020,Piotrowski2020,Piotrowski2020a,Qian2021,Rawson2019,Rawson2021,Richter2020a,SanchezStern2020,SanchezStern2023,Selsam2019,Selsam2019a,Si2020,Suda2021,Suda2021a,Wang2020a,Wang2021,Wang2021b,Wei2022,Xu2022a,Yan2023,Yang2019,You2019,Zhang2019,Zhang2020,Zhang2022,Zhu2022,Zombori2021,sun2020neurogift} & 70 \\
    RL & \cite{Abdelaziz2023,Brown2020a,Chen2019,Crouse2021,Fournier2022,Gauthier2020a,Gross2022,Hu2022,Jakubuuv2023,Kurin2020,Lassouaoui2019,Laurent2022,Lederman2020,Liu2023,Mo2020,Piepenbrock2021,Piepenbrock2022,Pimpalkhare2021,Poesia2021,Poesia2023,Scott2021,Shminke2022,Wang2020,Wang2023,Wu2020,Wu2021,Yolcu2019,Yu2023,Zheng2022,Zombori2019,Zombori2020,Zombori2021a} & 32 \\
    EA & \cite{Berden2022,Framil2022,Fu2020,Kumazawa2019,Kumazawa2021,Kumazawa2022,Kumazawa2023,Ma2019,Nagashima2019,Nawaz2021b,Nurcahyadi2022,Rafe2019,Sadeg2021,TsutomuKumazawa2022} & 14 \\
    NLP & \cite{Bordg2022,First2020,Han2021,Jiang2021,Jiang2022,Koscinski2021,Nayak2022,Palermo2022,Pimpalkhare2021a,Puccetti2021,Shigyo2020,Tworkowski2022,Urban2020,Welleck2022,Wu2022,Wu2022a} & 16 \\
\end{tabularx}

\clearpage
\begin{tabularx}{\textwidth}{@{}lXr@{}}
  \caption{Entries that use multiple AI types}%
  \label{tab:multp_ai_types} \\
  \toprule
  AI technique & Articles & Count \\
  \midrule
  \endfirsthead

  \caption{(Continued from previous page.)} \\
  \toprule
  AI technique & Articles & Count \\
  \midrule
  \endhead

  \bottomrule
  \multicolumn{3}{r@{}}{\small (Continued on next page.)} \\
  \endfoot

  \midrule
  {Unique total} & & 27 \\ 
  \bottomrule
  \endlastfoot

    RF & \cite{Atkari2020,Cai2019,Cai2019a,Dunkelau2020,Dunkelau2021,Lorenz2020,Nikolic2019,Scott2023,Zhang2021,Zhang2023,Zhu2019,Zhu2022a} & 12 \\
    SVM & \cite{Cai2019a,Cai2022,Dunkelau2021,PurgaL2021,Wang2019a,Zhu2022a} & 6 \\
    KNN & \cite{Dunkelau2021,Nikolic2019,Wang2019a,Zhang2021,Zhu2022a} & 5 \\
    NN & \cite{Atkari2020,Ayguen2022,Cai2019,Chvalovsky2019a,Dunkelau2019,Firoiu2021,Goertzel2021,Lorenz2020,Luo2023,Nawaz2021,PurgaL2021,Sadreddin2022,Scott2023} & 13 \\
    EA & \cite{Nawaz2021} & 1 \\
    Clustering & \cite{Holden2019,Liu2020b,Sadreddin2022} & 3 \\
    Data mining & \cite{Dunkelau2020} & 1 \\
    Logistic models & \cite{Atkari2020,Cai2019,Cai2019a,Cai2022,Nikolic2019,Zhu2022a} & 6 \\
    LR & \cite{Dunkelau2021,Nikolic2019,Wang2019a} & 3 \\
     & \cite{Zhu2019} & 1 \\
    Ridge regression & \cite{Dunkelau2021,Scott2023,Zhu2022a} & 3 \\
    Naive bayes & \cite{Atkari2020,Cai2019a,Cai2022,Zhu2022a} & 4 \\
    Trees & \cite{Atkari2020,Cai2019,Chvalovsky2019a,Dunkelau2021,Goertzel2021,Holden2019,Liu2020b,Scott2021a,Wang2019a,Zhang2021,Zhu2019,Zhu2022a} & 12 \\
    Nearest centroid & \cite{Zhu2022a} & 1 \\
    Passive aggressive & \cite{Zhu2022a} & 1 \\
    Stochastic gradient & \cite{Zhu2022a} & 1 \\
    LLM & \cite{Wang2023a,Zhang2023} & 2 \\
    Bagging & \cite{Wang2019a} & 1 \\
    Adaboost & \cite{Scott2021a,Wang2019a} & 2 \\
    Gaussian process & \cite{Zhu2022a} & 1 \\
    MCTS & \cite{Wang2023a} & 1 \\
    SMAC & \cite{Holden2019} & 1 \\
\end{tabularx}

\clearpage
\begin{tabularx}{\textwidth}{@{}llXr@{}}
  \caption{Entries grouped by AI types and contribution type}%
  \label{tab:ai_vs_contributions} \\
  \toprule
  Contribution type & AI technique & Articles & Count \\
  \midrule
  \endfirsthead

  \caption{(Continued from previous page.)} \\
  \toprule
  Contribution type & AI technique & Articles & Count \\
  \midrule
  \endhead

  \bottomrule
  \multicolumn{4}{r@{}}{\small (Continued on next page.)} \\
  \endfoot

  \midrule
  {Total} & & & 189 \\ 
  \bottomrule
  \endlastfoot

    Tool & KNN & \cite{Gauthier2020} & 1 \\
     & NN & \cite{Cherukuri2022,Huang2019,Laurent2021,Leeson2023,Nie2020,Rawson2019,Rawson2021,SanchezStern2020} & 8 \\
     & RL & \cite{Gross2022,Zombori2020,Zombori2021a} & 3 \\
     & NLP & \cite{First2020} & 1 \\
     & Mult.\ AI techniques & \cite{Scott2021a} & 1 \\
    Tool improvement & DT & \cite{Goertzel2019,Goertzel2020,Jakubuv2019} & 3 \\
     & NN & \cite{Chvalovsky2021,Chvalovskỳ2023,Suda2021a} & 3 \\
     & RL & \cite{Piepenbrock2022} & 1 \\
\end{tabularx}

\begin{tabularx}{\textwidth}{@{}llXr@{}}
  \caption{Entries grouped by AI types and publication years}%
  \label{tab:ai_vs_years} \\
  \toprule
  AI technique & Year & Articles & Count \\
  \midrule
  \endfirsthead

  \caption{(Continued from previous page.)} \\
  \toprule
  AI technique & Year & Articles & Count \\
  \midrule
  \endhead

  \bottomrule
  \multicolumn{4}{r@{}}{\small (Continued on next page.)} \\
  \endfoot

  \midrule
  {Total} & & & 189 \\ 
  \bottomrule
  \endlastfoot

    DT & 2019 & \cite{Blanchette2019,Goertzel2019,Jakubuv2019} & 3 \\
     & 2020 & \cite{Goertzel2020} & 1 \\
     & 2021 & \cite{Xu2021} & 1 \\
     & 2022 & \cite{Janota2022,Lample2022} & 2 \\
    RF & 2020 & \cite{Nejati2020} & 1 \\
     & 2022 & \cite{Beskyd2022} & 1 \\
    SVM & 2020 & \cite{Richter2020} & 1 \\
    KNN & 2020 & \cite{Blaauwbroek2020,Gauthier2020} & 2 \\
\end{tabularx}

\begin{tabularx}{\textwidth}{@{}lXr@{}}
  \caption{Entries grouped by FM types }%
  \label{tab:fm_types} \\
  \toprule
  FM technique & Articles & Count \\
  \midrule
  \endfirsthead

  \caption{(Continued from previous page.)} \\
  \toprule
  FM technique & Articles & Count \\
  \midrule
  \endhead

  \bottomrule
  \multicolumn{3}{r@{}}{\small (Continued on next page.)} \\
  \endfoot

  \midrule
  {Total} & & 189 \\ 
  \bottomrule
  \endlastfoot

    SAT & \cite{Amizadeh2019a,Atkari2020,Berden2022,Beskyd2022,Cameron2020,Chang2020,Chang2022,Dalla2021,Danisovszky2020a,Farooque2023,Fournier2022,Framil2022,Fu2020,Fuchs2023,Garzon2022,Han2020,Hireche2019,Hireche2020,Huang2022,Kumar2023,Kurin2020,Lassouaoui2019,Lederman2020,LeventiPeetz2020,Li2022,Lorenz2020,Marino2021,Nejati2020,Nurcahyadi2022,Ozolins2022,Richter2020,Sadeg2021,Sadreddin2022,Selsam2019,Selsam2019a,Wang2019a,Xu2021,Xu2022a,Yan2023,Yolcu2019,You2019,Zhang2020,Zhang2022,Zheng2022,sun2020neurogift} & 45 \\
    SMT & \cite{Blanchette2019,Dunkelau2019,Dunkelau2020,Hula2021,Jakubuuv2023,Janota2022,Leeson2023,Pimpalkhare2021,Pimpalkhare2021a,Scott2021,Scott2021a,Scott2023,Yao2021} & 13 \\
    TP & \cite{Abdelaziz2023,Ayguen2022,Baghdasaryan2021,Bansal2019,Bartek2020,Bartek2021,Blaauwbroek2020,Brown2020a,Chvalovsky2019a,Chvalovsky2021,Chvalovskỳ2023,Crouse2021,Faerber2020,Ferreira2020,Firoiu2021,First2020,First2022,Gauthier2020,Gauthier2020a,Glorot2019,Goertzel2019,Goertzel2019a,Goertzel2020,Goertzel2021,Goertzel2022,Han2021,Han2022,Holden2019,Huang2019,Jakubuv2019,Jakubuv2020,Jiang2021,Jiang2022,Kommrusch2023,Lample2022,Laurent2022,Liu2020b,Liu2022,Mangla2022,Mo2020,Morris2022,Nagashima2019,Nawaz2019,Nawaz2020,Nawaz2021,Nawaz2021b,Nie2020,Nikolic2019,Olsak2020,Palermo2022,Paliwal2020,Piepenbrock2021,Piepenbrock2022,Piotrowski2020,Piotrowski2020a,Poesia2021,Poesia2023,PurgaL2021,Qian2021,Rawson2019,Rawson2021,SanchezStern2020,SanchezStern2023,Shminke2022,Suda2021,Suda2021a,Tworkowski2022,Urban2020,Wang2020a,Wang2023a,Wei2022,Welleck2022,Wu2020,Wu2020a,Wu2021,Wu2022,Wu2022a,Yang2019,Zhang2019,Zhang2021,Zhang2023,Zombori2019,Zombori2020,Zombori2021,Zombori2021a} & 85 \\
    MC & \cite{Cao2022,Dunkelau2021,Gross2022,Hu2023,Jaeger2020,Kumazawa2019,Kumazawa2021,Kumazawa2023,Luo2023,Ma2019,Pira2022,Rafe2019,TsutomuKumazawa2022,Wang2020,Wang2021b,Zhu2019,Zhu2022,Zhu2022a} & 18 \\
\end{tabularx}

\begin{tabularx}{\textwidth}{@{}lXr@{}}
  \caption{Entries grouped by TP application types }%
  \label{tab:tp_types} \\
  \toprule
  Application goal & Articles & Count \\
  \midrule
  \endfirsthead

  \caption{(Continued from previous page.)} \\
  \toprule
  Application goal & Articles & Count \\
  \midrule
  \endhead

  \bottomrule
  \multicolumn{3}{r@{}}{\small (Continued on next page.)} \\
  \endfoot

  \midrule
  {Total} & & 85 \\ 
  \bottomrule
  \endlastfoot

    Heuristic selection & \cite{Holden2019,Nagashima2019} & 2 \\
    Premise selection & \cite{Bartek2020,Bartek2021,Chvalovsky2019a,Chvalovsky2021,Chvalovskỳ2023,Crouse2021,Ferreira2020,Firoiu2021,Goertzel2019,Goertzel2019a,Goertzel2021,Goertzel2022,Han2021,Jakubuv2019,Jakubuv2020,Jiang2022,Liu2020b,Liu2022,Mangla2022,Morris2022,Nawaz2021b,Piotrowski2020,Shminke2022,Suda2021,Tworkowski2022,Wei2022,Zombori2021} & 27 \\
    Proof search & \cite{Abdelaziz2023,Baghdasaryan2021,Bansal2019,Blaauwbroek2020,Faerber2020,Gauthier2020,Goertzel2020,Lample2022,Mo2020,Nawaz2021,Paliwal2020,Piepenbrock2021,Piepenbrock2022,Piotrowski2020a,Rawson2019,SanchezStern2020,Urban2020,Wang2023a,Zombori2019,Zombori2020,Zombori2021a} & 21 \\
    Proof mining & \cite{Jiang2021,Nawaz2020} & 2 \\
    Proof synthesis & \cite{Ayguen2022,First2020,First2022,Gauthier2020a,Glorot2019,Han2022,Kommrusch2023,Laurent2022,Palermo2022,Poesia2023,Qian2021,Rawson2021,SanchezStern2023,Wang2020a,Wu2020a,Wu2022,Wu2022a,Yang2019} & 18 \\
    Symbolic & \cite{Poesia2021} & 1 \\
    Formula synthesis & \cite{Brown2020a} & 1 \\
    Formula classification & \cite{PurgaL2021,Suda2021a} & 2 \\
    Tactic prediction & \cite{Nawaz2019,Wu2020,Zhang2019,Zhang2021} & 4 \\
    Lemma name generation & \cite{Nie2020} & 1 \\
    Portfolio solving & \cite{Nikolic2019} & 1 \\
    Mult.\ Techniques & \cite{Huang2019,Olsak2020,Welleck2022,Wu2021,Zhang2023} & 5 \\
\end{tabularx}

\begin{tabularx}{\textwidth}{@{}lXr@{}}
  \caption{Entries grouped by synthesis application types }%
  \label{tab:synthesis_types} \\
  \toprule
  Application goal & Articles & Count \\
  \midrule
  \endfirsthead

  \caption{(Continued from previous page.)} \\
  \toprule
  Application goal & Articles & Count \\
  \midrule
  \endhead

  \bottomrule
  \multicolumn{3}{r@{}}{\small (Continued on next page.)} \\
  \endfoot

  \midrule
  {Total} & & 19 \\ 
  \bottomrule
  \endlastfoot

    Invariant learning & \cite{Kobayashi2021,Liu2023,Wang2023} & 3 \\
    Loop invariant learning & \cite{Laurent2021,Xu2022b,Yao2020,Yu2023} & 4 \\
    Model repair & \cite{Cai2019,Cai2019a,Cai2022} & 3 \\
    Models/Specification & \cite{Abate2022,Bordg2022,Cherukuri2022,He2022,Hu2022,Koscinski2021,Nayak2022,Shigyo2020} & 8 \\
    Annotations & \cite{Puccetti2021} & 1 \\
\end{tabularx}

\begin{tabularx}{\textwidth}{@{}llXr@{}}
  \caption{Entries grouped by grade of TP automatization and TP application types  }%
  \label{tab:tp_auto_app_types} \\
  \toprule
  Application goal & Automation grade & Articles & Count \\
  \midrule
  \endfirsthead

  \caption{(Continued from previous page.)} \\
  \toprule
  Application goal & Automation grade & Articles & Count \\
  \midrule
  \endhead

  \bottomrule
  \multicolumn{4}{r@{}}{\small (Continued on next page.)} \\
  \endfoot

  \midrule
  {Total} & & & 85 \\ 
  \bottomrule
  \endlastfoot

    Heuristic selection & Automatic & \cite{Holden2019,Nagashima2019} & 2 \\
    Premise selection & Interactive & \cite{Jakubuv2019} & 1 \\
     & Automatic & \cite{Bartek2020,Bartek2021,Chvalovsky2019a,Chvalovsky2021,Chvalovskỳ2023,Crouse2021,Ferreira2020,Firoiu2021,Goertzel2019,Goertzel2019a,Goertzel2021,Goertzel2022,Han2021,Jakubuv2020,Jiang2022,Liu2020b,Liu2022,Mangla2022,Morris2022,Nawaz2021b,Piotrowski2020,Shminke2022,Suda2021,Tworkowski2022,Wei2022,Zombori2021} & 26 \\
    Proof search & Automatic & \cite{Abdelaziz2023,Baghdasaryan2021,Bansal2019,Blaauwbroek2020,Faerber2020,Gauthier2020,Goertzel2020,Lample2022,Mo2020,Nawaz2021,Paliwal2020,Piepenbrock2021,Piepenbrock2022,Piotrowski2020a,Rawson2019,SanchezStern2020,Urban2020,Wang2023a,Zombori2019,Zombori2020,Zombori2021a} & 21 \\
    Proof mining & Interactive & \cite{Nawaz2020} & 1 \\
     & Automatic & \cite{Jiang2021} & 1 \\
    Proof synthesis & Automatic & \cite{Ayguen2022,First2020,First2022,Gauthier2020a,Glorot2019,Han2022,Kommrusch2023,Laurent2022,Palermo2022,Poesia2023,Qian2021,Rawson2021,SanchezStern2023,Wang2020a,Wu2020a,Wu2022,Wu2022a,Yang2019} & 18 \\
    Symbolic & Automatic & \cite{Poesia2021} & 1 \\
\end{tabularx}

\clearpage
\begin{tabularx}{\textwidth}{@{}llXr@{}}
  \caption{Entries grouped by level of logic tackled by TP and TP application types  }%
  \label{tab:tp_logic_app_types} \\
  \toprule
  Application goal & Formalism & Articles & Count \\
  \midrule
  \endfirsthead

  \caption{(Continued from previous page.)} \\
  \toprule
  Application goal & Formalism & Articles & Count \\
  \midrule
  \endhead

  \bottomrule
  \multicolumn{4}{r@{}}{\small (Continued on next page.)} \\
  \endfoot

  \midrule
  {Total} & & & 85 \\ 
  \bottomrule
  \endlastfoot

    Heuristic selection & FOL & \cite{Holden2019} & 1 \\
     & HOL & \cite{Nagashima2019} & 1 \\
    Premise selection & FOL & \cite{Bartek2020,Bartek2021,Chvalovsky2019a,Chvalovsky2021,Chvalovskỳ2023,Crouse2021,Firoiu2021,Goertzel2019,Goertzel2019a,Goertzel2021,Goertzel2022,Jakubuv2019,Liu2020b,Liu2022,Mangla2022,Morris2022,Piotrowski2020,Shminke2022,Suda2021,Zombori2021} & 20 \\
     & HOL & \cite{Ferreira2020,Han2021,Jakubuv2020,Jiang2022,Nawaz2021b,Tworkowski2022,Wei2022} & 7 \\
    Proof search & FOL & \cite{Abdelaziz2023,Baghdasaryan2021,Faerber2020,Goertzel2020,Piotrowski2020a,Rawson2019,Urban2020,Zombori2020} & 8 \\
     & HOL & \cite{Bansal2019,Blaauwbroek2020,Gauthier2020,Lample2022,Mo2020,Nawaz2021,Paliwal2020,Piepenbrock2021,Piepenbrock2022,SanchezStern2020,Wang2023a,Zombori2019,Zombori2021a} & 13 \\
    Proof mining & HOL & \cite{Jiang2021,Nawaz2020} & 2 \\
    Proof synthesis & FOL & \cite{Ayguen2022,Glorot2019,Laurent2022,Palermo2022,Rawson2021,Wu2022,Wu2022a} & 7 \\
\end{tabularx}

\clearpage
\begin{tabularx}{\textwidth}{@{}llXr@{}}
  \caption{Entries grouped by FM techniques and years}%
  \label{tab:fm_years} \\
  \toprule
  FM technique & Year & Articles & Count \\
  \midrule
  \endfirsthead

  \caption{(Continued from previous page.)} \\
  \toprule
  FM technique & Year & Articles & Count \\
  \midrule
  \endhead

  \bottomrule
  \multicolumn{4}{r@{}}{\small (Continued on next page.)} \\
  \endfoot

  \midrule
  {Total} & & & 189 \\ 
  \bottomrule
  \endlastfoot

    SAT & 2019 & \cite{Amizadeh2019a,Hireche2019,Lassouaoui2019,Selsam2019,Selsam2019a,Wang2019a,Yolcu2019,You2019} & 8 \\
     & 2020 & \cite{Atkari2020,Cameron2020,Chang2020,Danisovszky2020a,Fu2020,Han2020,Hireche2020,Kurin2020,Lederman2020,LeventiPeetz2020,Lorenz2020,Nejati2020,Richter2020,Zhang2020,sun2020neurogift} & 15 \\
     & 2021 & \cite{Dalla2021,Marino2021,Sadeg2021,Xu2021} & 4 \\
     & 2022 & \cite{Berden2022,Beskyd2022,Chang2022,Fournier2022,Framil2022,Garzon2022,Huang2022,Li2022,Nurcahyadi2022,Ozolins2022,Sadreddin2022,Xu2022a,Zhang2022,Zheng2022} & 14 \\
     & 2023 & \cite{Farooque2023,Fuchs2023,Kumar2023,Yan2023} & 4 \\
    SMT & 2019 & \cite{Blanchette2019,Dunkelau2019} & 2 \\
     & 2020 & \cite{Dunkelau2020} & 1 \\
     & 2021 & \cite{Hula2021,Pimpalkhare2021,Pimpalkhare2021a,Scott2021,Scott2021a,Yao2021} & 6 \\
\end{tabularx}

\begin{tabularx}{\textwidth}{@{}llXr@{}}
  \caption{Entries grouped by FM techniques and contributions types}%
  \label{tab:fm_contributions} \\
  \toprule
  FM technique & Entry type & Articles & Count \\
  \midrule
  \endfirsthead

  \caption{(Continued from previous page.)} \\
  \toprule
  FM technique & Entry type & Articles & Count \\
  \midrule
  \endhead

  \bottomrule
  \multicolumn{4}{r@{}}{\small (Continued on next page.)} \\
  \endfoot

  \midrule
  {Total} & & & 189 \\ 
  \bottomrule
  \endlastfoot

    SAT & Workshop/short paper & \cite{Berden2022,Han2020,Yan2023} & 3 \\
     & Conference paper & \cite{Amizadeh2019a,Atkari2020,Beskyd2022,Cameron2020,Chang2020,Dalla2021,Danisovszky2020a,Farooque2023,Fournier2022,Framil2022,Garzon2022,Huang2022,Kurin2020,Lederman2020,LeventiPeetz2020,Li2022,Lorenz2020,Nejati2020,Ozolins2022,Sadeg2021,Sadreddin2022,Selsam2019,Selsam2019a,Wang2019a,Xu2021,Xu2022a,Yolcu2019,You2019,Zhang2020,Zhang2022,Zheng2022,sun2020neurogift} & 32 \\
     & Journal paper & \cite{Chang2022,Fu2020,Fuchs2023,Hireche2019,Hireche2020,Kumar2023,Lassouaoui2019,Marino2021,Nurcahyadi2022,Richter2020} & 10 \\
    SMT & Workshop/short paper & \cite{Blanchette2019,Jakubuuv2023} & 2 \\
     & Conference paper & \cite{Dunkelau2019,Dunkelau2020,Hula2021,Janota2022,Leeson2023,Pimpalkhare2021,Pimpalkhare2021a,Scott2021,Scott2021a,Yao2021} & 10 \\
     & Journal paper & \cite{Scott2023} & 1 \\
    TP & Workshop/short paper & \cite{Bartek2020,Brown2020a,Firoiu2021,Glorot2019,Goertzel2019a,Han2022,Holden2019,Liu2020b,Morris2022,Palermo2022,Piepenbrock2021,Qian2021,SanchezStern2020,Tworkowski2022,Wu2020,Wu2020a,Wu2022a,Zombori2019} & 18 \\
     & Conference paper & \cite{Ayguen2022,Bansal2019,Bartek2021,Blaauwbroek2020,Chvalovsky2019a,Chvalovsky2021,Chvalovskỳ2023,Crouse2021,Ferreira2020,First2022,Gauthier2020a,Goertzel2019,Goertzel2020,Goertzel2021,Goertzel2022,Han2021,Huang2019,Jakubuv2019,Jakubuv2020,Jiang2021,Jiang2022,Lample2022,Laurent2022,Mangla2022,Mo2020,Nagashima2019,Nawaz2019,Nie2020,Olsak2020,Paliwal2020,Piepenbrock2022,Piotrowski2020,Piotrowski2020a,Poesia2021,Rawson2019,Rawson2021,Suda2021,Suda2021a,Urban2020,Wang2020a,Wang2023a,Wei2022,Welleck2022,Wu2021,Wu2022,Yang2019,Zhang2019,Zhang2021,Zhang2023,Zombori2020,Zombori2021,Zombori2021a} & 52 \\
\end{tabularx}

\begin{tabularx}{\textwidth}{@{}llXr@{}}
  \caption{Entries grouped by FM and AI techniques}%
  \label{tab:fm_ai_contributions} \\
  \toprule
  FM technique & AI technique & Articles & Count \\
  \midrule
  \endfirsthead

  \caption{(Continued from previous page.)} \\
  \toprule
  FM technique & AI technique & Articles & Count \\
  \midrule
  \endhead

  \bottomrule
  \multicolumn{4}{r@{}}{\small (Continued on next page.)} \\
  \endfoot

  \midrule
  {Total} & & & 189 \\ 
  \bottomrule
  \endlastfoot

    SAT & DT & \cite{Xu2021} & 1 \\
     & RF & \cite{Beskyd2022,Nejati2020} & 2 \\
     & SVM & \cite{Richter2020} & 1 \\
     & NN & \cite{Cameron2020,Chang2020,Chang2022,Dalla2021,Danisovszky2020a,Farooque2023,Garzon2022,Han2020,Huang2022,LeventiPeetz2020,Li2022,Marino2021,Ozolins2022,Selsam2019,Selsam2019a,Xu2022a,Yan2023,You2019,Zhang2020,Zhang2022,sun2020neurogift} & 21 \\
     & RL & \cite{Fournier2022,Kurin2020,Lassouaoui2019,Lederman2020,Yolcu2019,Zheng2022} & 6 \\
     & EA & \cite{Berden2022,Framil2022,Fu2020,Nurcahyadi2022,Sadeg2021} & 5 \\
     & Clustering & \cite{Hireche2019,Hireche2020} & 2 \\
     & Mult.\ AI techniques & \cite{Atkari2020,Lorenz2020,Sadreddin2022,Wang2019a} & 4 \\
\end{tabularx}

\begin{tabularx}{\textwidth}{@{}llXr@{}}
  \caption{Entries grouped by TP and AI techniques}%
  \label{tab:tp_ai_contributions} \\
  \toprule
  Application goal & AI technique & Articles & Count \\
  \midrule
  \endfirsthead

  \caption{(Continued from previous page.)} \\
  \toprule
  Application goal & AI technique & Articles & Count \\
  \midrule
  \endhead

  \bottomrule
  \multicolumn{4}{r@{}}{\small (Continued on next page.)} \\
  \endfoot

  \midrule
  {Total} & & & 85 \\ 
  \bottomrule
  \endlastfoot

    Heuristic selection & EA & \cite{Nagashima2019} & 1 \\
     & Mult.\ AI techniques & \cite{Holden2019} & 1 \\
    Premise selection & DT & \cite{Goertzel2019,Jakubuv2019} & 2 \\
     & NN & \cite{Bartek2021,Chvalovsky2021,Chvalovskỳ2023,Ferreira2020,Goertzel2019a,Goertzel2022,Jakubuv2020,Liu2022,Piotrowski2020,Suda2021,Wei2022,Zombori2021} & 12 \\
     & RL & \cite{Crouse2021,Shminke2022} & 2 \\
     & EA & \cite{Nawaz2021b} & 1 \\
     & NLP & \cite{Han2021,Jiang2022,Tworkowski2022} & 3 \\
     & Bayesian inference & \cite{Mangla2022} & 1 \\
\end{tabularx}

\begin{tabularx}{\textwidth}{@{}llXr@{}}
  \caption{Entries grouped by SAT and AI techniques }%
  \label{tab:sat_ai_contributions} \\
  \toprule
  Application goal & AI technique & Articles & Count \\
  \midrule
  \endfirsthead

  \caption{(Continued from previous page.)} \\
  \toprule
  Application goal & AI technique & Articles & Count \\
  \midrule
  \endhead

  \bottomrule
  \multicolumn{4}{r@{}}{\small (Continued on next page.)} \\
  \endfoot

  \midrule
  {Unique total} & & & 45 \\ 
  \bottomrule
  \endlastfoot

    SAT solving & RF & \cite{Nejati2020} & 1 \\
     & NN & \cite{Selsam2019a,Xu2022a,Zhang2022,sun2020neurogift} & 4 \\
     & RL & \cite{Fournier2022,Yolcu2019} & 2 \\
     & Mult.\ AI techniques & \cite{Atkari2020} & 1 \\
     & Custom & \cite{Amizadeh2019a} & 1 \\
    Predict solvability & DT & \cite{Xu2021} & 1 \\
     & NN & \cite{Cameron2020,Chang2022,Selsam2019,Zhang2020} & 4 \\
    MaxSat & NN & \cite{Chang2020,Marino2021} & 2 \\
\end{tabularx}

\begin{tabularx}{\textwidth}{@{}llXr@{}}
  \caption{Entries grouped by synthesis and AI techniques }%
  \label{tab:synth_ai_contributions} \\
  \toprule
  Application goal & AI technique & Articles & Count \\
  \midrule
  \endfirsthead

  \caption{(Continued from previous page.)} \\
  \toprule
  Application goal & AI technique & Articles & Count \\
  \midrule
  \endhead

  \bottomrule
  \multicolumn{4}{r@{}}{\small (Continued on next page.)} \\
  \endfoot

  \midrule
  {Total} & & & 19 \\ 
  \bottomrule
  \endlastfoot

    Invariant learning & NN & \cite{Kobayashi2021} & 1 \\
     & RL & \cite{Liu2023,Wang2023} & 2 \\
    Loop invariant learning & NN & \cite{Laurent2021} & 1 \\
     & RL & \cite{Yu2023} & 1 \\
     & Clustering & \cite{Xu2022b} & 1 \\
     & Custom & \cite{Yao2020} & 1 \\
    Model repair & Mult.\ AI techniques & \cite{Cai2019,Cai2019a,Cai2022} & 3 \\
    Models/Specification & NN & \cite{Cherukuri2022,He2022} & 2 \\
     & RL & \cite{Hu2022} & 1 \\
     & NLP & \cite{Bordg2022,Koscinski2021,Nayak2022,Shigyo2020} & 4 \\
     & Custom & \cite{Abate2022} & 1 \\
    Annotations & NLP & \cite{Puccetti2021} & 1 \\
\end{tabularx}

\fi

\end{appendices}

\section*{Declarations}

\bmhead*{Funding.}
The authors have no relevant financial or non-financial interests to disclose.

\bmhead*{Ethical approval.}
Not applicable.

\bmhead*{Informed consent.}
Not applicable.

\bmhead*{Author Contributions.}
The first author conceived the idea of conducting the systematic mapping study.
The first and second authors conducted the search process outlined in
\cref{sec:strategy} together.
The application of IC/EC was done independently by the first and second authors.
Disagreements in this filtering process were solved by the third author.
The first and second authors were responsible for the manuscript's writing.
The third author reviewed the manuscript, provided feedback, and supervised
the project. The public repository containing the dataset of found and investigated studies
was prepared by the second author.

\bmhead*{Data Availability Statement.}
The primary studies found in this systematic mapping study are publicly available
at \url{https://github.com/hhu-stups/ai4fm-studies}.

\bmhead*{Conflict of Interest.}
The authors have no competing interests to declare that are relevant to the
content of this article.

\bmhead*{Clinical Trial Number.}
Not applicable.

\clearpage
\addcontentsline{toc}{section}{References}
\bibliography{search_results,not_in_time_frame,fm-ml-mapping.bib}

\end{document}